\documentclass[preprint,12pt]{elsarticle}

\usepackage{amssymb}
\usepackage{amsmath}

\usepackage{graphicx}
\usepackage{float}

\usepackage[ruled, vlined, linesnumbered]{algorithm2e}
\DontPrintSemicolon%
\usepackage{multirow}
\usepackage{makecell}
\usepackage{multicol}
\usepackage{booktabs}
\floatstyle{plaintop}
\usepackage{subcaption}

\usepackage[newfloat,frozencache,cachedir=.]{minted}
\newenvironment{code}{\floatstyle{plaintop}%
\captionsetup{type=listing, labelfont=bf,justification=raggedright, singlelinecheck=false}}{}
\SetupFloatingEnvironment{listing}{name=\textbf{Listing}}

\makeatletter
\newcommand*\mysize{%
  \@setfontsize\mysize{8}{8}%
}
\makeatother

\makeatletter
\def\ps@pprintTitle{%
  \let\@oddhead\@empty
  \let\@evenhead\@empty
  \let\@oddfoot\@empty
  \let\@evenfoot\@oddfoot
}
\makeatother

\setminted{fontsize=\mysize}

\usepackage{enumitem}

\newcommand{\benchmark}[0]{\textsc{C-Pack-IPAs}\xspace}
\newcommand{\tcas}[0]{\textsc{TCAS}\xspace}
\newcommand{\iscas}[0]{\textsc{ISCAS85}\xspace}
\newcommand{\cfaults}[0]{\textsc{CFaults}\xspace}
\newcommand{\bugassist}[0]{\textsc{Bug\-Assist}\xspace}
\newcommand{\sniper}[0]{\textsc{SNI\-PER}\xspace}
\newcommand{\hsd}[0]{\textsc{HSD}\xspace}
\newcommand{\cbmc}[0]{\textsc{CBMC}\xspace}

\newtheorem{definition}{Definition}
\newtheorem{example}{Example}

\usepackage{xcolor}
\definecolor{pastelblue}{RGB}{80,140,200}
\definecolor{pastelgreen}{RGB}{80,170,120}
\definecolor{pastelpurple}{RGB}{140,120,200}
\usepackage[
    colorlinks=true,
    linkcolor=pastelblue,
    citecolor=pastelgreen,
    urlcolor=pastelpurple
]{hyperref}

\begin{document}

\begin{frontmatter}



\title{Model-Based Diagnosis with Multiple Observations: A Unified Approach for C Software and Boolean Circuits} 

\author[label1]{Pedro Orvalho\footnote{PO is now affiliated with IIIA-CSIC, Spain. The work was conducted while he was at the University of Oxford.}
}
\affiliation[label1]{organization={Department of Computer Science, University of Oxford},
            city={Oxford},
            country={United Kingdom}}

\author[label1]{Marta Kwiatkowska}

\author[label2]{Mikoláš Janota}
\affiliation[label2]{organization={Czech Technical University in Prague},
            city={Prague},
            country={Czech Republic}}

\author[label3]{Vasco Manquinho}
\affiliation[label3]{organization={INESC-ID, Instituto Superior Técnico, Universidade de Lisboa},
            city={Lisbon},
            country={Portugal}}

\begin{abstract}
Debugging is one of the most time-consuming and expensive tasks in software development and circuit design. Several formula-based fault localisation~(FBFL) methods have been proposed, but they fail to guarantee a set of diagnoses across all failing tests or may produce redundant diagnoses that are not subset-minimal, particularly for programs/circuits with multiple faults.

This paper introduces \cfaults, a novel fault localisation tool for C software and Boolean circuits with multiple faults. \cfaults leverages Model-Based Diagnosis (MBD) with multiple observations and aggregates all failing test cases into a unified Maximum Satisfiability (MaxSAT) formula. Consequently, our method guarantees consistency across observations and simplifies the fault localisation procedure. 
Experimental results on three benchmark sets, two of C programs, \tcas and \benchmark, and one of Boolean circuits, \iscas, show that \cfaults is faster at localising faults in C software than other FBFL approaches such as \bugassist, \sniper, and \hsd. 
On the \iscas benchmark, \cfaults is generally slower than \hsd; however, it localises faults in only 6\% fewer circuits, demonstrating that it remains competitive in this domain. 
Furthermore, \cfaults produces only subset-minimal diagnoses of faulty statements, whereas the other approaches tend to enumerate redundant diagnoses~(e.g.,~\bugassist~and~\sniper).
\vspace{-0.05in}
\end{abstract}








\end{frontmatter}



\section{Introduction}
\label{sec:intro}

Localising system faults has long been recognised as one of the most time-consuming and expensive tasks in Artificial Intelligence~(AI)~\cite{reiter87,DBLP:journals/ai/KleerW87,DBLP:conf/aaai/MetodiSKC12,bugAssist-cav11,CFaults-FM24}. 
Given a buggy system, \emph{fault localisation~(FL)} is the process of identifying the locations in the system that may cause faulty behaviour~(bugs)~\cite{reiter87}. 
More specifically, given a faulty system and a set of observations (i.e., failing test cases), \emph{formula-based fault localisation (FBFL)} methods encode the localisation problem as a series of constraint optimisation problems to identify a minimal set of faulty statements (diagnoses) within the system. 
These FBFL approaches leverage the theory of \emph{Model-Based Diagnosis (MBD)}~\cite{reiter87}, which has been applied to restore consistency across several domains, including Boolean circuits~\cite{DBLP:journals/ai/KleerW87,DBLP:conf/aaai/MetodiSKC12,DBLP:conf/aaai/SternKFP12,DBLP:journals/jair/MetodiSKC14,ijcai15-Marques-SilvaJI15,ijcai19-ignatievMWM}, C software~\cite{bugAssist-pldi11,jip16-SNIPER,CFaults-FM24}, knowledge bases~\cite{DBLP:journals/ws/ShchekotykhinFFR12,DBLP:conf/dx/RodlerS17,DBLP:journals/kbs/Rodler22}, and spreadsheets~\cite{jannach2010toward,DBLP:journals/ai/RodlerHJNW25}. 
Typically, FBFL methods compute a minimal diagnosis by considering each failing test case individually rather than simultaneously across all failing test cases. 
Moreover, most of these methods~\cite{bugAssist-cav11,bugAssist-pldi11,ijcai15-Marques-SilvaJI15,jip16-SNIPER,ijcai19-ignatievMWM,CFaults-FM24} usually enumerate all \emph{Minimal Correction Subsets~(MCSes)}~\cite{DBLP:journals/jar/LiffitonS08} of a \emph{Maximum Satisfiability}~(MaxSAT) formula to cover all possible diagnoses, and then apply different aggregation mechanisms to determine the minimal diagnosis that explains the system inconsistency.

For instance, \bugassist~\cite{bugAssist-pldi11,bugAssist-cav11}, a prominent FBFL tool for C software, implements a ranking mechanism for bug locations. As explained in Section~\ref{sec:bugassist}, for each failing test, \bugassist enumerates all diagnoses of a MaxSAT formula corresponding to bug locations. Subsequently, \bugassist ranks diagnoses based on their frequency of appearance in each failing test. 
Other FBFL tools, like \sniper~\cite{jip16-SNIPER}, also enumerate all diagnoses for each failing test. However, as explained in Section~\ref{sec:sniper}, the set of \sniper's diagnoses is obtained by taking the Cartesian product of the diagnoses gathered using each failing test.
As a result, while FBFL methods can determine minimal diagnoses per failing test, \bugassist cannot guarantee a minimal diagnosis considering all failing tests, and \sniper may enumerate a significant number of redundant diagnoses that are not minimal~\cite{ijcai19-ignatievMWM}. These limitations may pose challenges for C programs with multiple faulty statements, as shown~in~Example~\ref{eg:motivation}.

\begin{table*}[t]
\begin{minipage}[t!]{0.47\columnwidth}
\centering
\begin{code}
\caption{Faulty program example. Faulty lines: \{5,8,11\}.}
\label{code:motivating_eg}
\begin{minted}[escapeinside=||,tabsize=0,obeytabs,xleftmargin=1.5em,linenos]{C}
int main(){
  // finds maximum of 3 numbers
  int f,s,t;
  scanf("%d%d%d",&f,&s,&t);
  if (f < s && f >= t)
    // fix: f >= s
    printf("%d",f);
  if (f > s && s <= t)
    // fix: f < s and s >= t
    printf("%d",s);
  if (f > t && s > t)
    // fix: f < t and s < t
    printf("%d",t);

  return 0;
}
\end{minted}
\end{code}
\end{minipage}
\begin{minipage}[t!]{0.5\columnwidth}
\centering
\resizebox{0.85\columnwidth}{!}{%
\begin{tabular}{|l|*{3}{wc{1cm}|}l|wc{1.5cm}|}
\cline{2-4} \cline{6-6}
\multicolumn{1}{c|}{} & \multicolumn{3}{c|}{\textbf{Input}} &  & \textbf{Output} \\ \cline{1-4} \cline{6-6}
{\textbf{$t_0$}} & 1 & 2 & 3 &  & 3 \\ \cline{1-4} \cline{6-6}
{\textbf{$t_1$}} & 6 & 2 & 1 &  & 6 \\ \cline{1-4} \cline{6-6}
{\textbf{$t_2$}} & -1 & 3 & 1 &  & 3 \\ \cline{1-4} \cline{6-6}
\end{tabular}
}
\centering
\caption{Test-suite.}
\label{tab:test-suite}
\hfill
\resizebox{0.85\columnwidth}{!}{%
\centering
\begin{tabular}{c|c|c|}
\cline{2-3}
& \textbf{\bugassist}                 & \textbf{\sniper}            \\ \hline
\multicolumn{1}{|c|}{\textbf{\#Diagnoses $t_0$}}  & 8   &  8\\ \hline
\multicolumn{1}{|c|}{\textbf{\#Diagnoses $t_1$}}  &  21  & 21  \\ \hline
\multicolumn{1}{|c|}{\textbf{\#Diagnoses $t_2$}}  & 9  & 9 \\ \hline      
\multicolumn{1}{|c|}{\textbf{\begin{tabular}[c]{@{}c@{}}\#Total\\Unique Diagnoses\end{tabular}}} & 32 & 1297 \\ \hline
\multicolumn{1}{|c|}{\textbf{Final Diagnosis}} & \{4,13\} & \{5,8,11\} \\ \hline
\end{tabular}%
}
\caption{Number of diagnoses (faulty statements) generated by \bugassist~\cite{bugAssist-cav11} and \sniper~\cite{jip16-SNIPER} per test.}
\label{tab:diagnoses}

\end{minipage}
\end{table*}

\begin{example}[Motivation]
\label{eg:motivation}
Consider the C program presented in Listing~\ref{code:motivating_eg}, which aims to determine the maximum among three given numbers. However, based on the test suite shown in Table~\ref{tab:test-suite}, the program is faulty, as its output differs from the expected. The set of minimally faulty lines in this program is \{5, 8, 11\}, as all three {\tt if}-conditions are incorrect according to the test suite. Fixing any subset of these lines would be insufficient to repair the program. One possible fix is to replace all these conditions with the suggested fixes in lines \{6, 9, 12\}.

In a typical FBFL approach for C software, the minimal set of statements identified as faulty might include, for example, lines 4 and 5. Removing the \texttt{scanf} statement and an {\tt if}-statement would allow an FBFL tool to assign any value to the input variables in order to always produce the expected output.
However, considering an approach that prioritises identifying faulty statements within the program's logic before evaluating issues in the input/output statements (such as \texttt{scanf} and \texttt{printf}), one might identify lines \{5, 8, 11\} as the faulty statements.
When applying \bugassist's and \sniper's approach on the program in Listing~\ref{code:motivating_eg} with the described optimisation criterion and utilising the inputs/outputs detailed in Table~\ref{tab:test-suite} as specification, distinct sets of faults are identified for each failing test. Table~\ref{tab:diagnoses} presents the diagnosis (set of faulty lines) produced by each tool, along with the number of diagnoses enumerated for each failing test case and the total number of unique diagnoses after aggregating the diagnoses from all tests, using each tool's respective method.

In the case of \bugassist, diagnoses are prioritised based on their occurrence frequency. Consequently, \bugassist yields 32 unique diagnoses and selects \{4, 13\} since this diagnosis is identified in every failing test. In contrast, \sniper computes the Cartesian product of all diagnoses, resulting in 1297 unique diagnoses.
Note that \bugassist's diagnoses may not adequately identify all faulty program statements. Conversely, \sniper's final diagnosis \{5, 8, 11\} is minimal, even though it enumerates an additional 1296 diagnoses.
Hence, existing FBFL methods do not ensure a minimal diagnosis across all failing tests (e.g., \bugassist) or may produce an overwhelming number of redundant sets of diagnoses (e.g., \sniper), especially for programs with multiple faults.
\end{example}

This paper tackles this challenge by formulating the fault localisation (FL) problem as a single optimisation problem, as described in Section~\ref{sec:mbd-multiple-tests}. 
We leverage MaxSAT and the theory of \emph{Model-Based Diagnosis (MBD)}~\cite{reiter87,ijcai19-ignatievMWM,DBLP:journals/ai/Rodler23}, integrating all failing test cases simultaneously. 
This approach enables the generation of only minimal diagnoses that identify all faulty components within a system, in our case either a C program or a Boolean circuit. 
Furthermore, we implement the MBD problem with multiple test cases in \cfaults, a fault localisation tool designed to \emph{see faults in C} programs and Boolean \emph{C}ircuits, presented in Section~\ref{sec:cfaults}. 

For C programs, \cfaults first unrolls and instruments the code at the program level, ensuring independence from the bounded model checker. 
It then calls \cbmc~\cite{tacas04-cbmc-ClarkeKL}, a well-known bounded model checker for C, to generate a trace formula. 
Finally, \cfaults encodes the problem into MaxSAT to identify the minimal set of diagnoses corresponding to~the~buggy~statements. 

For Boolean circuits, \cfaults first unrolls and instruments the given circuit for each failing observation. 
The resulting unrolled and instrumented circuit is then directly translated into a unified MaxSAT formula. 
A MaxSAT solver is invoked to compute the minimal diagnosis that restores consistency between the Boolean circuit and all observations. 

In addition, we adapt the \emph{hitting
set dualisation} (\hsd)~\cite{ijcai19-ignatievMWM} algorithm, the current publicly available state-of-the-art FBFL method for Boolean circuits, to perform fault localisation in C software.

Experimental results presented in Section~\ref{sec:results-c}, on two benchmarks of C programs, \tcas~\cite{tcas-dataset-ESE05} (industrial) and \benchmark~\cite{C-Pack-IPAs_apr24} (programming exercises), show that \cfaults consistently detects only minimal sets of diagnoses and outperforms other FBFL methods such as \bugassist, \sniper, and \hsd in terms of speed. 
In contrast, \sniper and \bugassist either generate an overwhelming number of redundant diagnoses or fail to produce the minimal sets required to repair each program. 
Moreover, \cfaults identifies faults in 31\% more programs than \bugassist, 36\% more programs than \sniper, and 15\% more programs than \hsd. 
Section~\ref{sec:results-circuits} further presents experiments using \cfaults and \hsd to localise bugs in Boolean circuits on the \iscas~\cite{iscas85} benchmark, showing that while \cfaults is generally slower than \hsd, it localises only 6\% fewer faulty circuits, thereby remaining competitive for Boolean circuit fault localisation.
Finally, Section~\ref{sec:use-cases} presents several use cases where \cfaults has been successfully applied, Section~\ref{sec:related} reviews related work, and the paper concludes in Section~\ref{sec:conclusion}.

To summarise, the contributions of this work are: 
\begin{itemize}
\item We address the fault localisation problem in C programs and Boolean circuits using a Model-Based Diagnosis (MBD) approach that considers multiple failing test cases and formulates the problem as a unified optimisation task.

\item Our MBD approach will be made publicly available in an upcoming release of \cfaults~\cite{CFaults-Zenodo-FM2024}\footnote{\url{https://github.com/pmorvalho/CFaults}}, our fault localisation tool that unrolls and instruments C programs at the code level, thereby decoupling it from any specific bounded model checker. That release will also support fault localisation in Boolean circuits.

\item Experiments using \cfaults on two sets of C programs (\tcas and \benchmark) show that \cfaults is faster, and localises faults in more programs than the other localisation approaches.

\item \cfaults produces only subset-minimal diagnoses, unlike other state-of-the-art formula-based fault localisation (FBFL) tools, such as \bugassist and \sniper.

\item On \iscas, a benchmark of Boolean circuits, \cfaults is generally slower than \hsd.  However, it localises faults in only 6\% fewer circuits, demonstrating that it remains competitive in this domain.
\end{itemize}

\paragraph{Contributions Beyond the Conference Version}
This article substantially extends the conference paper
\emph{``CFaults: Model-Based Diagnosis for Fault Localisation in C with Multiple Test Cases''}~\cite{CFaults-FM24}
by introducing new theoretical developments, improved techniques, and a broader experimental evaluation.
\emph{First}, the original method, designed exclusively for C software, is \emph{generalised} to also localise faults in Boolean circuits, providing a single formula-based framework capable of handling both C programs and Boolean circuits under multiple failing observations.
\emph{Second}, several relaxation mechanisms for program statements are revised to correctly address specific issues, such as non-terminating loops, and uninitialised global variables, thus ensuring complete fault localisation under our model-based diagnosis (MBD) approach (see Section~\ref{sec:cfaults}).
\emph{Third}, we include a detailed empirical comparison with \hsd~\cite{ijcai19-ignatievMWM}, the publicly available state-of-the-art formula-based fault-localisation method for Boolean circuits, highlighting the strengths and limitations of our unified approach.
\emph{Fourth}, the background and theoretical exposition are significantly expanded to offer a more comprehensive discussion of MBD, including additional explanations of core algorithms and related approaches, such as \bugassist, \sniper, and \hsd.
\emph{Finally}, the experimental evaluation is significantly extended: the assessment on C software now covers the full \benchmark benchmark~\cite{C-Pack-IPAs_apr24}, increasing from roughly 500 to nearly 1500 higher-complexity programs; additionally, we report new experiments on Boolean circuits using the \iscas benchmark~\cite{iscas85}.

\section{Preliminaries}
\label{sec:prelim}

This section provides definitions that are used throughout the paper.

\begin{definition}[Boolean Satisfiability (SAT)]
The Boolean Satisfiability (SAT) problem is the decision problem for propositional logic~\cite{biere2009handbook}. 
\end{definition}

A propositional formula in Conjunctive Normal Form (CNF) is a conjunction of clauses where each clause is a disjunction of literals.
A literal is a propositional variable $x_i$ or its negation $\neg x_i$.
Given a CNF formula $\phi$, the SAT problem corresponds to deciding if there is an assignment to the variables in $\phi$ such that $\phi$ is satisfied or prove that no such assignment exists.
When applicable, set notation will be used for formulas and clauses. A formula can be represented as a set of clauses (meaning its conjunction) and a clause as a set of literals (meaning its disjunction).

\begin{example}[Satisfiable Formula]
  Let $\phi$ denote a CNF formula with the following clauses:
  $\{ (x_1), (\neg x_1 \vee x_2), (x_2 \vee x_3), (\neg x_1 \vee \neg x_3) \}$.
  In this case, the assignment $\{ (x_1, 1), (x_2, 1), (x_3, 0) \}$ satisfies $\phi$.
  \label{ex:sat}
\end{example}

\begin{example}[Unsatisfiable Formula]
  Let $\phi$ denote a CNF formula as follows:
  $\{ (x_1), (\neg x_1 \vee x_2), (x_2 \vee x_3), (\neg x_1 \vee \neg x_3), (x_3) \}$.
  Since there is no assignment to the variables such that $\phi$ is satisfied,  we say that $\phi$~is~unsatisfiable.
  \label{ex:unsat}
\end{example}

\begin{definition}[SAT Solver Call]
Given a CNF formula $\phi$, the result of the SAT solver call $\text{SAT}(\phi)$ is a triple $(st, \nu, \phi_C)$, where $st$ denotes the solver status (\texttt{SAT} or \texttt{UNSAT}). 
If the call is \texttt{SAT}, $\nu$ is a model of $\phi$.
Otherwise, $\phi_C \subseteq \phi$ contains an \texttt{unsatisfiable core}, that is, a subformula for which no assignment exists that satisfies all its clauses simultaneously~\cite{DBLP:conf/cp/SiZMJIN16}.
\end{definition}

\begin{example}[Unsatisfiable Core]
  Let $\phi$ denote a CNF formula as follows:
  $\{ (x_1), (\neg x_1 \vee x_2), (x_2 \vee x_3), (\neg x_1 \vee \neg x_3), (x_3) \}$. $\phi$~is~unsatisfiable, and $\phi_C = \{ (x_1), (\neg x_1 \vee \neg x_3), (x_3) \}$ is an unsatisfiable core
  since there is no assignment that satisfies all the clauses in $\phi_C$ simultaneously.
  \label{ex:unsat-core}
\end{example}

\begin{definition}[Maximum Satisfiability (MaxSAT)]
The Maximum Satisfiability (MaxSAT) problem is an optimisation version of the SAT problem. Given a CNF formula $\phi$, the goal is to find an assignment that maximises the number of satisfied clauses in $\phi$~\cite{li2009maxsat,handbook-maxsat}. 
\end{definition}

In partial MaxSAT, $\phi$ is split into hard clauses ($\phi_h$) and soft clauses ($\phi_s$). Given a formula $\phi = (\phi_h, \phi_s)$, the goal is to find an assignment that satisfies all hard clauses in $\phi_h$ while minimising the number of unsatisfied soft clauses in $\phi_s$. Moreover, in the weighted version of the partial MaxSAT problem, each soft clause is assigned a weight, and the goal is to find an assignment that satisfies all hard clauses and minimises the sum of the weights of the unsatisfied soft clauses. 

\begin{example}[MaxSAT]
  Consider the partial MaxSAT formula $\phi = (\phi_h, \phi_s)$ where $\phi_h = \{ (x_1 \vee x_2), (\neg x_2 \vee x_3) \}$ and $\phi_s = \{ (\neg x_1), (\neg x_3) \}$. In this case, the assignment $\{ (x_1, 1), (x_2, 0), (x_3, 0) \}$ is an optimal assignment to $\phi$ since it only unsatisfies one clause in $\phi_s$ while satisfying all clauses in $\phi_h$. 
\end{example}

Several MaxSAT algorithms rely on iterative calls to a SAT solver. In particular, \emph{Core-guided MaxSAT algorithms}~\cite{DBLP:conf/cp/SiZMJIN16,DBLP:journals/jsat/MorgadoIM14,DBLP:conf/cp/MorgadoDM14,DBLP:conf/aaai/HerasMM11} have proven highly effective for instances arising from real-world applications~\cite{DBLP:journals/jsat/MorgadoIM14}. 
These algorithms exploit the ability of SAT solvers to identify \emph{unsatisfiable~cores}. 

\begin{definition}[Minimal Correction Subset (MCS)]
Given a MaxSAT formula $\phi = (\phi_h, \phi_s)$, a Minimal Correction Subset (MCS) $\kappa$ of $\phi$ is a subset $\kappa \subseteq \phi_s$ where $\phi_h \cup (\phi_s \setminus \kappa)$ is satisfiable and, for all $c \in \kappa$, $\phi_h \cup (\phi_s \setminus \kappa) \cup \lbrace c \rbrace$ is unsatisfiable.    
\end{definition}

\begin{example}[MCS]
  Consider the partial MaxSAT formula $\phi = (\phi_h, \phi_s)$ where $\phi_h = \{ (x_1 \vee x_2), (x_2 \vee \neg x_3), (\neg x_2 \vee x_3) \}$ and $\phi_s = \{ (\neg x_1), (\neg x_2), (\neg x_3) \}$. 
  One possible MCS would be $C_1 = \{ (\neg x_1) \}$. Note that $\phi_h \cup \phi_s$ is unsatisfiable, but $\phi_h \cup \phi_s \setminus C_1$ is satisfiable and $C_1$ is minimal. Another MCS of the formula is $C_2 = \{ (\neg x_2), (\neg x_3) \}$.
  \label{ex:mcs}
\end{example}

A dual concept of MCSes are \emph{Minimal Unsatisfiable Subsets~(MUSes)}~\cite{DBLP:journals/jar/LiffitonS08,DBLP:conf/ijcai/Marques-SilvaHJPB13}.

\begin{definition}[Minimal Unsatisfiable Subset (MUS)]
Let $\phi = (\phi_h, \phi_s)$ denote a MaxSAT formula. A Minimal Unsatisfiable Subset (MUS) $\mu$ of $\phi$ is a subset $\mu \subseteq \phi_s$ where $\phi_h \cup \mu$ is unsatisfiable and, for all $c \in \mu$, $\phi_h \cup \mu \setminus \{ c\}$ is satisfiable.    
\end{definition}

\begin{example}[MUS]
  Consider the partial MaxSAT formula $\phi = (\phi_h, \phi_s)$ where $\phi_h = \{ (x_1 \vee x_2), (x_2 \vee \neg x_3), (\neg x_2 \vee x_3) \}$ and $\phi_s = \{ (\neg x_1), (\neg x_2), (\neg x_3) \}$. 
  One possible MUS would be $\mu_1 = \{ (\neg x_1), (\neg x_2) \}$. Note that $\phi_h \cup \mu_1$ is unsatisfiable and $\mu_1$ is minimal. Another MUS is $\mu_2 = \{ (\neg x_1), (\neg x_3) \}$.
  \label{ex:mus}
\end{example}

\begin{definition}[Program]
A program is considered sequential, comprising standard statements such as assignments, conditionals, loops, and function calls, each adhering to their conventional semantics in C. A program is deemed to contain a bug when an assertion violation occurs during its execution with input $I$. Conversely, if no assertion violation occurs, the program is considered correct for input~$I$. In cases where a bug is detected for input $I$, it is possible to define an error trace, representing the sequence of statements executed by program $P$ on input $I$~\cite{tacas04-cbmc-ClarkeKL}.
\end{definition}

\begin{definition}[Trace Formula (TF)]
A Trace Formula (TF) is a propositional formula that is SAT iff there exists an execution of the program that terminates with a violation of an assert statement while satisfying all assume~statements~\cite{bugAssist-pldi11}.    
\end{definition}

For further information on TFs, interested readers are referred to~\cite{tacas04-cbmc-ClarkeKL,DBLP:conf/dac/ClarkeKY03}.

\section{Model-Based Diagnosis With One Observation}
\label{sec:mbd}

The following definitions are commonly used in the \emph{Model-Based Diagnosis (MBD)} theory~\cite{ijcai19-ignatievMWM,reiter87,ijcai15-Marques-SilvaJI15}. 
A system description $\mathcal{P}$ is composed of a set of components $\mathcal{C} = \{c_1, \ldots, c_n\}$. Each component in $\mathcal{C}$ can be declared \emph{healthy} or \emph{unhealthy}. For each component $c \in \mathcal{C}$, $h(c) = 0$ if $c$ is unhealthy, otherwise, $h(c) = 1$. As in prior works~\cite{ijcai19-ignatievMWM,DBLP:journals/jair/MetodiSKC14}, $\mathcal{P}$ is described by a CNF formula, where $\mathcal{F}_c$ denotes the encoding of component~$c$:

\begin{equation}
    \mathcal{P} \triangleq \bigwedge\nolimits_{c \in \mathcal{C}} { ( \neg h(c) \vee \mathcal{F}_c )}
  \label{eq:system-description}
\end{equation}

Observations represent deviations from the expected system behaviour. An observation, denoted as $o$, is a finite set of first-order sentences~\cite{reiter87,ijcai19-ignatievMWM}, which is assumed to be encodable in CNF as a set of unit clauses. In this work, the failing test cases represent the set of observations.

A system $\mathcal{P}$ is considered faulty if there exists an inconsistency with a given observation $o$ when all components are declared healthy. The problem of model-based diagnosis (MBD) aims to identify a set of components which, if declared unhealthy, restore consistency. This problem is represented by the 3-tuple $\langle \mathcal{P}, \mathcal{C}, o\rangle$, and can be encoded as a CNF formula:

\begin{equation}
    \mathcal{P} \wedge o \wedge \bigwedge\nolimits_{c \in \mathcal{C}} { h(c) } \vDash \bot
  \label{eq:mbd}
\end{equation}

For a given MBD problem $\langle \mathcal{P}, \mathcal{C}, o\rangle$, a set of system components $\Delta \subseteq \mathcal{C}$ is a diagnosis iff:
    \begin{equation}
    \mathcal{P} \wedge o \wedge \bigwedge\nolimits_{c \in \mathcal{C} \setminus \Delta } { h(c) }  \wedge \bigwedge\nolimits_{c \in \Delta} { \neg h(c) } \nvDash \bot
  \label{eq:diagnosis}
  \end{equation}
  
A diagnosis $\Delta$ is minimal iff no subset of $\Delta$, $\Delta' \subsetneq \Delta$, is a diagnosis, and $\Delta$ is of minimal cardinality if there is no other diagnosis $\Delta'' \subseteq \mathcal{C}$ with $|\Delta''| < |\Delta|$.

A diagnosis is redundant if it is not subset-minimal~\cite{ijcai19-ignatievMWM}. 

To encode the Model-Based Diagnosis problem with one observation with partial MaxSAT, the set of clauses that encode $\mathcal{P}$ (\ref{eq:system-description}), and the set of clauses that encode observation $o$, represent the set of hard clauses. The soft clauses consist of unit clauses that aim to maximise the set of healthy components, i.e., $\bigwedge_{c \in \mathcal{C}} { h(c) }$~\cite{DBLP:conf/fmcad/SafarpourMVLS07,ijcai15-Marques-SilvaJI15}. This MaxSAT encoding of MBD enables enumerating minimum cardinality diagnoses and subset minimal diagnoses, considering a single observation. 
Furthermore, a minimal diagnosis is a minimal correction subset~(MCS) of the MaxSAT formula. Given an inconsistent formula that encodes the MDB problem (\ref{eq:mbd}), a minimal diagnosis $\Delta$ satisfies (\ref{eq:diagnosis}), thereby making $\Delta$ an MCS of the MaxSAT formula. \bugassist~\cite{bugAssist-pldi11}, \sniper~\cite{jip16-SNIPER}, and other model-based diagnosis~(MBD) tools for fault localisation in circuits~\cite{ijcai15-Marques-SilvaJI15,DBLP:conf/fmcad/SafarpourMVLS07} encode with partial MaxSAT the localisation problem using a single failing observation.

\begin{algorithm}[t!]
\caption{\bugassist's Fault Localisation Algorithm~\cite{bugAssist-pldi11}.}
\label{alg:bugassist}
\KwIn{Program $P$, assertion $p$, failing tests $\{t_1, \dots, t_k\}$}
\KwOut{Ranked list of unique MCSes based on frequency across failing tests}

$\mathcal{M}_{all} \gets []$ \tcp*{List of MCS sets for each test}

\ForEach{test $t_i \in \{t_1, \dots, t_k\}$}{
    $(\texttt{test}, \sigma) \gets \textsf{GenerateTrace}(P, t_i)$\;
    \If{$\sigma = \texttt{None}$}{
        \textbf{continue} \tcp*{Skip if no trace was found}
    }

    $\Phi_H \gets \texttt{test} \land p \land \textit{TF}_1(\sigma)$\;
    $\Phi_S \gets \textit{TF}_2(\sigma)$\;
    $\mathcal{M}_i \gets EnumerateMCSes(\Phi_H, \Phi_S)$ \tcp*{MCSes for this test}
    $\mathcal{M}_{all} \gets \mathcal{M}_{all} \cup \{\mathcal{M}_i\}$\;
}

$\mathcal{R} \gets \textsf{VoteMCS}(\mathcal{M}_{all})$\ \label{vote-mcses};

\Return $\mathcal{R}$

\end{algorithm}

\subsection{\bugassist}
\label{sec:bugassist}

\bugassist~\cite{bugAssist-cav11,bugAssist-pldi11}, an FBFL approach for C software, prioritises diagnoses according to their frequency of occurrence. 
Algorithm~\ref{alg:bugassist} outlines \bugassist's localisation procedure. 
For each failing test case, \bugassist encodes the MBD problem, using a single observation (i.e., a failing test case), as a MaxSAT formula following the encoding in (\ref{eq:mbd}), and enumerates all minimal correction subsets~(MCSes). 
It then invokes Algorithm~\ref{alg:bugassist-vote-mcses} to rank each MCS based on its frequency across the failing test cases. 
The MCS that occurs most frequently and contains the fewest faulty components is ultimately returned to the user.

\begin{algorithm}[t!]
\caption{VoteMCS: Aggregate and Rank MCSes by Frequency and Size}
\label{alg:bugassist-vote-mcses}
\KwIn{$t$ sets of MCSes: $\mathcal{M}_1, \mathcal{M}_2, \dots, \mathcal{M}_t$}
\KwOut{Sorted list $\mathcal{R}$ of unique MCSes}

$\mathcal{F} \gets \emptyset$ \tcp*{Map: MCS $\to$ frequency count}

\ForEach{$\mathcal{M}_i$ in $\{\mathcal{M}_1, \dots, \mathcal{M}_t\}$}{
    \ForEach{MCS $S$ in $\mathcal{M}_i$}{
        \If{$S \notin \mathcal{F}$}{
            $\mathcal{F}[S] \gets 1$\;
        }
        \Else{
            $\mathcal{F}[S] \gets \mathcal{F}[S] + 1$\;
        }
    }
}

$\mathcal{R} \gets \textsf{Keys}(\mathcal{F})$\;
Sort $\mathcal{R}$ by: (1) descending frequency $\mathcal{F}[S]$; and then (2) ascending set size $|S|$ to break ties;

\Return $\mathcal{R}$

\end{algorithm}

\begin{algorithm}[t!]
\caption{\sniper's SetCombine Function~\cite{jip16-SNIPER}.}
\label{alg:sniper-combine}
\KwIn{A set of sets of MCSes $D = \{D_1, D_2, \dots, D_n\}$}
\KwOut{Set of complete diagnoses $\mathcal{C}$}

$\mathcal{C} \gets \emptyset$\;

\ForEach{tuple $(d_1, d_2, \dots, d_n) \in D_1 \times D_2 \times \dots \times D_n$}{
    $C \gets \bigcup_{i=1}^{n} d_i$\;
    $\mathcal{C} \gets \mathcal{C} \cup \{C\}$\;
}

\Return $\mathcal{C}$
\end{algorithm}

\subsection{\sniper}
\label{sec:sniper}

Similarly to \bugassist, \sniper also enumerates all MCSes for each failing test case. 
The key difference between the two FBFL methods lies in how they aggregate this information to determine the most appropriate MCS to return to the user (line~\ref{vote-mcses} in Algorithm~\ref{alg:bugassist}). 
Unlike \bugassist, which treats each failing test case independently, \sniper performs a Cartesian product over all generated MCSes, combining information across different failing test cases using Algorithm~\ref{alg:sniper-combine}. 
This strategy allows \sniper to compute a \emph{subset-minimal aggregated diagnosis}, i.e., the smallest set of faulty components that explains all failing test cases. 
However, achieving this diagnosis requires \sniper to enumerate a large number of redundant diagnoses, which significantly increases computational overhead.

\section{Model-Based Diagnosis with Multiple Observations}
\label{sec:mbd-multiple-tests}

More recently, the MaxSAT encoding for MBD has been generalised to multiple inconsistent observations~\cite{ijcai19-ignatievMWM}.
Let $\mathcal{O} = \{o_1,\ \dots\ o_m\}$ be a set of observations. Each observation is associated with a replica $\mathcal{P}_i$ of the system~$\mathcal{P}$. The system remains unchanged given different observations, where the components are replicated for each observation, but the healthy variables are shared. For a given observation $o_i$, a diagnosis is given by the following:
    \begin{equation}
    \mathcal{P}_i \wedge o_i \wedge \bigwedge\nolimits_{c \in \mathcal{C} \setminus \Delta } { h(c) }  \wedge \bigwedge\nolimits_{c \in \Delta} { \neg h(c) } \nvDash \bot
   \label{eq:diagnosis-multiple}
   \end{equation}

The goal is to find a minimal diagnosis $\Delta \subseteq \mathcal{C}$, such that $\Delta$ is a minimal set of components when deactivated the system becomes consistent with all observations $\mathcal{O} = \{o_1,\ \dots\ o_m\}$. Moreover, when considering multiple observations, an aggregated diagnosis is a subset of components that includes one possible diagnosis for each given observation.

\subsection{\hsd}
\label{sec:hsd}

\begin{algorithm}[t!]
\caption{Hitting Set Dualisation Fault (\hsd) Localisation Algorithm with Multiple Observations~\cite{ijcai19-ignatievMWM}.}
\label{alg:hsd}
\KwIn{System description \texttt{P}, Observations $O\ =\ \{o_1, \dots, o_m$\}}
\KwOut{All subset-minimal aggregated diagnoses $\mathcal{D} = \{\Delta_1, \Delta_2, \ldots\}$ and explanations $\mathcal{U} = \{U_1, U_2, \ldots\}$}

$(P_1, \dots, P_m) \gets \textsf{Encode}(\texttt{P}, \texttt{o}_1, \dots, \texttt{o}_m)$ \tcp*{$m$ replicas of $P$ and observations, encoded into $m$ formulae (hard clauses)}
$\phi_S \gets \{h(c) \mid c \in \mathcal{C}\}$ \tcp*{Healthy variables (soft clauses)}
$\mathcal{D} \gets \emptyset$ \tcp*{Diagnoses}

\While{true}{
    $(\texttt{status}, \Delta) \gets 
    \textsf{MinHS}(\mathcal{U}, \mathcal{D})$ 
    \tcp*{Minimal HS of diagnoses}
    \If{\texttt{status} = \texttt{false}}{
        \textbf{break}\;
    }

    \ForEach{$i \in \{1, \dots, m\}$}{
        $(\texttt{status}_i, \kappa) \gets \textsf{SAT} (P_i \cup (\phi_S \setminus \Delta))$\; \label{line:core-computation}
        \If{\texttt{status}\_i = \texttt{false}}{
            $U \gets \textsf{Reduce}(\kappa)$ \tcp*{Extract MUS (explanation)} \label{line:reduce}
            $\mathcal{U} \gets \mathcal{U} \cup \{U\}$\;
            \textbf{break loop}\;
        }
    }

    \If{all $P_i \cup (\phi_S \setminus \Delta)$ are satisfiable}{
        $\mathcal{D} \gets \mathcal{D} \cup \{\Delta\}$\;
    }

    \ForEach{$i \in \{1, \dots, m\}$}{
        \If{\textsf{SAT}$(P_i \cup \mathcal{D}) = \texttt{false}$}{
                \Return\;
            }
    }

}
\end{algorithm}

The HSD algorithm~\cite{ijcai19-ignatievMWM}, presented in Algorithm~\ref{alg:hsd}, was proposed for fault localisation in systems based on multiple observations. It is based on \emph{hitting set dualisation (HSD)}~\cite{DBLP:conf/aaai/SternKFP12,DBLP:conf/cp/DaviesB11,DBLP:conf/soda/ChandrasekaranKMV11}. For each observation $o_i$, the algorithm computes minimal unsatisfiable subsets (MUSes) from the MaxSAT encoding defined in Equation~(\ref{eq:diagnosis-multiple}). It then calculates a minimum hitting set $\Delta$ over the unsatisfiable cores (MUSes if core minimisation is applie) and checks whether $\Delta$ makes the system consistent with each observation individually. To compute all subset-minimal aggregated diagnoses for a faulty system $\mathcal{P}$, the algorithm performs at most $m$ oracle calls for each minimum hitting set, where $m$ is the number of observations. Each oracle call uses a distinct system replica based on Equation~(\ref{eq:diagnosis-multiple}).

Note that, in line~\ref{line:core-computation}, if the minimum hitting set $\Delta$ of the MUSes is not an aggregated diagnosis, i.e., it does not constitute a diagnosis for at least one of the observations, the SAT oracle returns \texttt{UNSAT} together with an unsatisfiable core $\kappa$ of the formula (see Section~\ref{sec:prelim}). 
The \hsd algorithm can be run in two configurations: one that applies minimisation to the \emph{unsatisfiable cores} $\Delta$ in line~\ref{line:reduce}, referred to as \hsd-CM, and another that does not perform this minimisation, referred to simply as \hsd. The core minimisation step consists of at least one additional oracle call that allows the algorithm to reduce the working formula~$\mathcal{U}$. 
This helps to minimise memory usage, although a potential drawback is the increase in computation time caused by the extra(s) SAT call(s). 
Note that the number of cores computed by this algorithm corresponds directly to the number of iterations performed by the \hsd algorithm.

Moreover, we would like to point out that \hsd has only been evaluated on Boolean circuits with single faults, and its publicly available implementation does not support FBFL for C programs.

\subsection{Our approach}
\label{sec:our-approach}
\begin{algorithm}[t!]
\caption{Enumerate All MaxSAT Solutions}
\label{alg:cfaults}
\KwIn{System description $P$, Observations $O\ =\ \{o_1, \dots, o_m$\}}
\KwOut{All subset-minimal aggregated diagnoses $\mathcal{D} = \{\Delta_1, \Delta_2, \dots\}$}

$\mathcal{D} \gets \emptyset$ \tcp*{Diagnoses}
$\texttt{optCost} \gets \texttt{None}$ \tcp*{Best known cost}

$\phi_H \gets \bigcup_{i=1}^m P_i \vee o_i$ \tcp*{$m$ replicas of $P$ and observations, encoded into one formula (hard clauses)}
$\phi_S \gets \{h(c) \mid c \in \mathcal{C}\}$ \tcp*{Soft clauses (healthy variables)}

\While{true}{
    $(\Delta, \texttt{cost}) \gets \textsf{SolveMaxSAT}(\phi_H, \phi_S)$\;
    \If{$\Delta = \texttt{None}$}{
        \textbf{break} \tcp*{No more solutions}
    }

    \If{$\texttt{optCost} = \texttt{None}$}{
        $\texttt{optCost} \gets \texttt{cost}$ \tcp*{Initialise optimal cost}
    }

    \If{$\texttt{cost} > \texttt{optCost}$}{
        \textbf{break} \tcp*{All MaxSAT solutions enumerated.}
    }

    $\mathcal{D} \gets \mathcal{D} \cup \{\Delta\}$\;

    Add blocking clause: $\bigvee_{h(c) \in \Delta} \neg h(c)$ to $\phi_H$\;
}

\Return $\mathcal{D}$
\end{algorithm}

This paper encodes the fault localisation problem as a Model-Based Diagnosis with multiple observations using a single optimisation problem. We simultaneously integrate all failing test cases (observations) in a single MaxSAT formula. This approach allows us to generate only minimal diagnoses capable of identifying all faulty components within the system, in our case, a C program or a Boolean circuit. Algorithm~\ref{alg:cfaults} presents our approach.

Given $m$ observations, $\mathcal{O} = \{o_1, \dots, o_m\}$, a distinct replica of the system, denoted as $\mathcal{P}_i$, is required for each observation $o_i$. The hard clauses, $\phi_h$, in our MaxSAT formulation correspond to each observation's encoding ($o_i$) and $m$ system replicas, one for each observation, $\mathcal{P}_i$. Hence, $\phi_h = \bigwedge_{o_i \in \mathcal{O}} {( \mathcal{P}_i \wedge o_i )}$. Additionally, we aim to maximise the set of healthy components. Therefore, the soft clauses are formulated as: $\phi_s = \bigwedge_{c \in \mathcal{C} } { h(c) }$. Thus, given the MaxSAT solution of $(\phi_h, \phi_s)$, the set of unhealthy components  ($h(c)=0$), corresponds to a subset-minimal aggregated diagnosis, which also represents the smallest minimal diagnosis. This diagnosis is a subset-minimal of components that, when declared unhealthy (deactivated), make the system consistent with all observations, as follows:

\begin{equation}     
     \bigwedge\nolimits_{o_i \in \mathcal{O}} {( \mathcal{P}_i \wedge o_i )} \wedge \bigwedge\nolimits_{c \in \mathcal{C} \setminus \Delta } { h(c) }  \wedge \bigwedge\nolimits_{c \in \Delta} { \neg h(c) } \nvDash \bot
  \label{eq:diagnosis-multiple-single-formula}
\end{equation}

We assume that the system remains unchanged given different observations, where the components are replicated for each observation, but the healthy variables are shared. This is necessary because we analise all observations jointly, which can affect the component's behaviour.

\subsection{Our Approach vs \hsd}
Our approach encodes the problem into a single MaxSAT formula, whereas \hsd~\cite{ijcai19-ignatievMWM} divides it into $m$ MaxSAT formulas, one for each observation. 
Moreover, for each minimum hitting set computed by HSD, $m$ oracle calls are required to verify whether a diagnosis is consistent with all observations. 
In contrast, our method requires only a single MaxSAT call, which directly returns a cardinality minimal diagnosis that is, by definition, consistent with all observations, since they are all encoded within the same formula. 
Furthermore, the \hsd algorithm has been evaluated solely on circuits with single faults under multiple observations and was not originally implemented for programs. 
We have adapted \hsd to localise faults in C software, and in Section~\ref{sec:results} we compare the performance of \cfaults against \hsd on both C programs and Boolean circuits.

\section{\cfaults: MBD with Multiple Observations for C Software and Boolean Circuits}%
\label{sec:cfaults}

\begin{figure}[t!]
    \centering
    \includegraphics[width=\columnwidth]{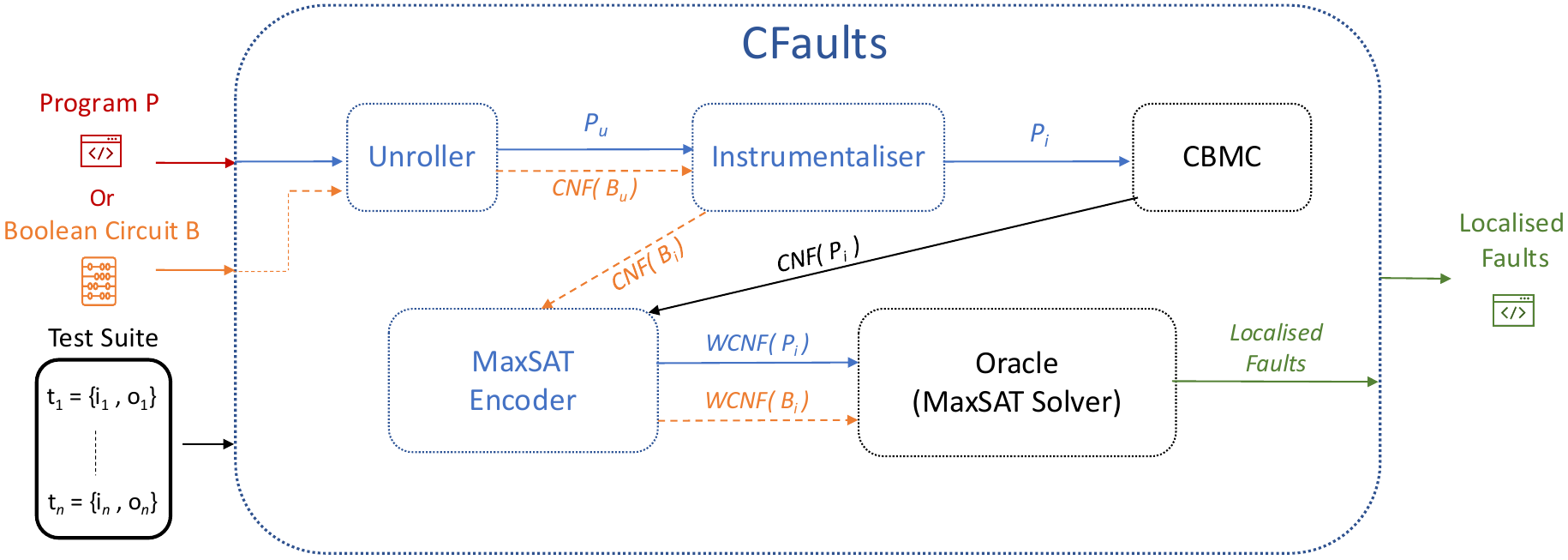}
    \caption{Overview of \cfaults.}
    \label{fig:cfaults-overview}
\end{figure}

This section presents \cfaults, which is a model-based diagnosis (MBD) tool for fault localisation in C programs and Boolean circuits with multiple failing test cases.
Unlike previous works, \cfaults uses the approach proposed in Section~\ref{sec:our-approach}, and C programs are relaxed at the code level, enabling users to leverage other bounded model checkers effectively.
Figure~\ref{fig:cfaults-overview} provides an overview of \cfaults consisting of five steps: system unrolling, system instrumentalisation, bounded model checking (\cbmc), encoding to MaxSAT, and an Oracle~(MaxSAT~solver).

As explained in Section~\ref{sec:cfaults-progs} and illustrated by the blue arrows in Figure~\ref{fig:cfaults-overview}, when \cfaults receives a C program and a test suite as input, all processing steps are executed. In this case, \cfaults formulates the MBD problem with multiple test cases as the 3-tuple $\langle \mathcal{P}_i, \mathcal{C}, \mathcal{O} \rangle$, where the observations $\mathcal{O}$ correspond to failing test cases (inputs and assertions), the components $\mathcal{C}$ denote the set of program statements, and the system description $\mathcal{P}_i$ is a trace formula derived from the unrolled and instrumented program. The program is instrumented at the code level using relaxation variables, referred to as \emph{healthy variables}.

Furthermore, as detailed in Section~\ref{sec:cfaults-circuits} and indicated by the orange dashed arrows in Figure~\ref{fig:cfaults-overview}, when \cfaults receives a Boolean circuit, \texttt{B}, and a test suite as input, only four steps are performed: unrolling the circuit, instrumenting the unrolled circuit, encoding it into MaxSAT, and calling an oracle. In this scenario, \cfaults formulates the MBD problem as the 3-tuple $\langle B_i, \mathcal{C}, \mathcal{O} \rangle$, where the observations $\mathcal{O}$ consist of failing input–output test cases, the components $\mathcal{C}$ represent the set of gates in the Boolean circuit, and the system description $B_i$ is a CNF formula encoding the unrolled and instrumentalised Boolean circuit \texttt{B}.

\subsection{\cfaults sees Faults in C Programs}
\label{sec:cfaults-progs}

Before unrolling the given C program, \cfaults first runs the static analysers \textsc{cppcheck}~\cite{cppcheck} and \textsc{clang-tidy}~\cite{clang-tidy} to detect issues, such as, uninitialised variables and potential runtime errors (e.g., division by zero). If any issues are detected, \cfaults immediately reports them to the user; otherwise, it proceeds to unroll the buggy program. We integrated static analysers after our user study~\cite{ojm24-sigcse} showed their practical value in helping to pinpoint these problems in students’ programming assignments (see~Section~\ref{sec:use-cases}).

\subsubsection{Program unrolling} 
\cfaults starts the unrolling process by expanding the faulty program using the set of failing tests from the test suite. In this context, an unrolled program signifies the original program expanded $m$ times ($m$ program scopes), where $m$ denotes the number of failed test cases. An unrolled program encodes the execution of all failing tests within the program, along with their corresponding inputs and specifications (assertions).

The unrolling process encompasses three primary steps. Initially, \cfaults generates \emph{fresh variables} and functions for each of the $m$ program scopes, ensuring each scope possesses unique variables and functions. Subsequently, \cfaults establishes variables representing the inputs and outputs for each program scope corresponding to the failing tests. 

Input operations, such as \texttt{scanf}, undergo translation into read accesses to arrays corresponding to the inputs, while output operations, such as \texttt{printf}, are replaced by write operations into arrays representing the program's output. Every exit point of the program (e.g., a {\tt return} statement in the {\tt main} function) is replaced with a \texttt{goto} statement directing the program flow to the next failing test's scope. Furthermore, to avoid the C compiler’s default zero-initialisation, all uninitialised global variables in the program are explicitly initialised to nondeterministic values. 

Lastly, at the end of the unrolled program, \cfaults embeds an assertion capturing all the specifications of the failing tests.
Thus, the unrolled program encapsulates the execution of all failing tests within a single program. 

Listing~\ref{code:prog_unrolled} exhibits a program segment generated through the unrolling process applied to Listing~\ref{code:motivating_eg}. \cfaults establishes global variables to represent the inputs and outputs of each failing test (lines 1--3, Listing~\ref{code:prog_unrolled}). 
For the sake of simplicity, the depicted listing illustrates solely the initial scope corresponding to test 0 from the test suite outlined in Table~\ref{tab:test-suite}. Distinct variables are introduced for each failing test. Furthermore, the \texttt{scanf} function call is substituted with input array operations (lines 8--10), while the \texttt{printf} calls are replaced with \cfaults' print functions, akin to \texttt{sprintf} functions, which direct output to a buffer. Lastly, the unrolled program concludes with an assertion representing the disjunction of the negation of all failing test assertions. For instance, suppose there are $m$ failing tests, where $A_i$ denotes the assertion of test $t_i$. In this scenario, \cfaults injects the following assertion into the program: $\neg A_1 \lor \dots \lor \neg A_m$. 

\begin{figure*}[t]
\begin{code}
\captionof{listing}{The program from Listing~\ref{code:motivating_eg} after being subjected to \cfaults' unrolling process, using the test suite presented in Table~\ref{tab:test-suite}. For simplicity, only the initial scope corresponding to test $t_0$ is displayed. The scopes \texttt{scope\_1} and \texttt{scope\_2} associated with failing tests $t_1$ and $t_2$ are omitted.}
\label{code:prog_unrolled}
\begin{minted}[tabsize=0,obeytabs,xleftmargin=20pt,linenos]{C}
float _input_f0[3] = {1, 2, 3};
char _out_0[2] = "3";
int _ioff_f0 = 0, _ooff_0 = 0;
// ... inputs and outputs for the other tests
int main(){
  scope_0:{
    int f_0, s_0, t_0;
    f_0 = _input_f0[_ioff_f0++];
    s_0 = _input_f0[_ioff_f0++];
    t_0 = _input_f0[_ioff_f0++];
    if ((f_0 < s_0) && (f_0 >= t_0))
        _ooff_0 = printInt(_out_0, _ooff_0, f_0);
    if ((f_0 > s_0) && (s_0 <= t_0))
        _ooff_0 = printInt(_out_0, _ooff_0, s_0);
    if ((f_0 > t_0) && (s_0 > t_0))
        _ooff_0 = printInt(_out_0, _ooff_0, t_0);
    goto scope_1;    
  }
  // ... scope_1 and scope_2
  final_step:
  assert(strcmp(_out_0, "3") != 0 || // other assertions);
}
\end{minted}
\end{code}
\vspace{-.5cm}
\end{figure*}

\subsubsection{Program Intrumentalisation} 
\label{sec:c-instrumentalisation}
After integrating all possible executions and assertions from failing tests during the unrolling step, \cfaults proceeds to instrumentalise the unrolled C program by introducing \emph{relaxation variables} for each program component~(i.e., statement, instruction). These relaxation variables correspond to the healthy variables explained in Section~\ref{sec:mbd-multiple-tests}. Thus, each relaxation variable activates (or deactivates) the program component being relaxed when assigned to \texttt{true} (or \texttt{false}) respectively. \cfaults ensures that there are no conflicts between the names of the relaxation variables and the names of the program's original variables. For this step, \cfaults needs to receive a maximum number of iterations that the program should be unwound.

The relaxation process introduces relaxation variables that deactivate or activate program components. This process involves four distinct relaxation rules for: (1) conditions of {\tt if}-statements, (2) expression lists (e.g., an expression list executed at the beginning of a for-loop), (3) loop conditions, and (4) other program statements.

\begin{example}
\label{eg:code-relaxation}
Listings~\ref{code:prog_statements} shows a code snippet that sums all numbers between $1$ and \texttt{n}. Listings~\ref{code:statements_relaxed} depicts the same program statements after undergoing relaxation by \cfaults. For the sake of simplicity, all relaxation variable and offset names were simplified.
\end{example}

\begin{table*}[t]
\begin{minipage}[t!]{0.45\columnwidth}
\begin{code}
\vspace{-.88in}
\captionof{listing}{Program statements.}
\label{code:prog_statements}
\begin{minted}[escapeinside=||,tabsize=0,obeytabs,xleftmargin=20pt,linenos]{C}
int i;
int n;
int s;

s = 0;
n = _input_f0[_ioff_f0++];

if (n == 0)
    return 0;

for (i=1; i < n; i++){
    s = s + i;
}
\end{minted}
\end{code}
\end{minipage}
\begin{minipage}[t!]{0.54\columnwidth}
\begin{code}
\captionof{listing}{Program statements relaxed.}
\label{code:statements_relaxed}
\begin{minted}[escapeinside=çç,tabsize=1,obeytabs,xleftmargin=20pt,linenos]{C}
//main scope
bool ç\textcolor{blue}{\textit{\_rv1, \_rv2, \_rv3, \_rv5}}ç;
bool ç\textcolor{blue}{\textit{\_rv6[UNWIND],..., \_rv9[UNWIND]}}ç;
int ç\textcolor{blue}{\textit{\_los}}ç; // loop1 offset

//test scope 
bool ç\textcolor{blue}{\textit{\_ev4, \_ev7[UNWIND]}}ç;
int i,n,s;
ç\textcolor{blue}{\textit{\_los=1}}ç;

if (ç\textcolor{blue}{\textit{\_rv1}}ç) s = 0;
if (ç\textcolor{blue}{\textit{\_rv2}}ç) n = _input_f0[_ioff_f0++];

if ( ç\textcolor{blue}{\textit{\_rv3}}ç ? (n == 0) : ç\textcolor{blue}{\textit{\_ev4}}ç)
    return 0;

for (ç\textcolor{blue}{\textit{\_rv5}}ç ? (i = 1) : ç\textcolor{blue}{\textit{1}}ç; 
     ç\textcolor{blue}{\textit{\_rv6[\_los]}}ç ? (i<n) : ç\textcolor{blue}{\textit{\_ev7[\_los]}}ç; 
     ç\textcolor{blue}{\textit{\_rv9[\_los]}}ç ? i++ : ç\textcolor{blue}{\textit{1, \_los++}}ç){
    if (ç\textcolor{blue}{\textit{\_rv8[\_los]}}ç) s = s + i;       
}
\end{minted}
\end{code}
\end{minipage}
\vspace{-0.15in}
\end{table*}

In more detail, the rule for relaxing a general program statement is to envelop the statement with an {\tt if}-statement, whose condition is a relaxation variable. For example, consider lines 5 and 6 in the program on Listings~\ref{code:prog_statements}. These lines are relaxed by \cfaults using relaxation variables \texttt{\_rv1} and \texttt{\_rv2} respectively, appearing as lines 11 and 12 on Listings~\ref{code:statements_relaxed}.

Furthermore, when relaxing {\tt if}-statements, the statements inside the {\tt then} and {\tt else} blocks adhere to the previously explained relaxation rule. However, the conditions of {\tt if}-statements are relaxed using a ternary operator, as shown in line 14 of Listings~\ref{code:statements_relaxed}. Note that if the relaxation variable is assigned \texttt{true}, then the original {\tt if} condition is executed. Otherwise, a different relaxation variable (e.g., \texttt{\_ev4} in Listings~\ref{code:statements_relaxed}) determines whether the program execution enters the {\tt then}-block or the {\tt else}-block (if one exists).
These relaxation variables (\emph{{\tt else}'s relaxation variables}) are local to each failing test scope and enable different tests to determine whether to enter the {\tt then} or {\tt else}-block.

When handling expression lists, \cfaults adopts a comparable strategy to that of generic program statements, enclosing each expression within a ternary operator instead of an {\tt if}-statement. If the program component is deactivated, the expression is replaced by \texttt{1}. For example, the initialisation of variable \texttt{i} in line 11 of Listings~\ref{code:prog_statements} is relaxed into the ternary operation in line 17 of Listings~\ref{code:statements_relaxed}.

Lastly, all relaxation variables inside a loop are Boolean vectors to relax statements within the loop. Each entry of these vectors relaxes the loop's statements for a given iteration. The maximum number of iterations of the loops is defined by the \cfaults user.
\cfaults follows a similar approach for inner loops, creating arrays of arrays. Thus, for simple program statements within a loop, \cfaults encapsulates them with {\tt if}-statements, with the relaxation variables indexed to the iteration number. Line 20 of Listings~\ref{code:statements_relaxed} illustrates a relaxed statement inside a loop. The loop's condition is relaxed by implication of the relaxation variable, as demonstrated in line 18 of Listings~\ref{code:statements_relaxed}. Furthermore, each loop has its own offsets to index relaxation variables. These offsets are initialised just before the loop and incremented at the end of each iteration (e.g., line 19 in Listing~\ref{code:statements_relaxed}).

We would like to point out that our relaxation of loop conditions (see line 18 in Listing~\ref{code:statements_relaxed}) differs from the approach adopted in the previous version of this work~\cite{CFaults-FM24}, which relaxed the loop condition as follows: 
\begin{center}
\texttt{(\textcolor{blue}{\textit{!\_rv6[\_los]}} || (i<n))}    
\end{center}
However, this relaxation strategy was incomplete, as it did not account for non-terminating loops, i.e., scenarios where the loop condition is always assigned to true. In contrast, our current approach mirrors the relaxation method applied to {\tt if}-statement conditions, and we introduce \emph{{\tt else} relaxation variables} to control the execution path when the original condition is deactivated. This ensures that the analysis accurately captures all control flow behaviours, including potentially non-terminating loops.

When handling auxiliary functions, \cfaults declares the relaxation variables needed in the main scope of the program and passes these variables as parameters. Hence, \cfaults ensures that the same variables are used throughout the auxiliary functions' calls.

Listing~\ref{code:prog_inst} depicts the program resulting from the instrumentalisation process of Listing~\ref{code:prog_unrolled} performed by \cfaults. 
The same program components (i.e., statements, instructions) across different failing test scopes are assigned the same relaxation variable declared in the main scope. Consequently, if a relaxation variable is set to \texttt{false}, the corresponding program component is deactivated across all test executions.  Additionally, the relaxation variables are left uninitialised, allowing \cfaults to determine the minimum number of faulty components requiring deactivation. Note that relaxation variables are not declared as global variables but as local variables within the {\tt main} scope. This is to prevent the C compiler from automatically initialising all these variables to \texttt{false}.

\begin{figure*}[t]
\begin{code}
\captionof{listing}{Instrumentalised program.}
\label{code:prog_inst}
\begin{minted}[escapeinside=çç,tabsize=1,obeytabs,xleftmargin=20pt,linenos]{C}
//global vars
int main(){
  bool ç\textcolor{blue}{\textit{\_rv1}}ç, ç\textcolor{blue}{\textit{\_rv2}}ç, ..., ç\textcolor{blue}{\textit{\_rv12}}ç;
  scope_0:{
    bool ç\textcolor{blue}{\textit{\_ev5}}ç, ç\textcolor{blue}{\textit{\_ev8}}ç, ç\textcolor{blue}{\textit{\_ev11}}ç;
    int f_0, s_0, t_0;
    if (ç\textcolor{blue}{\textit{\_rv1}}ç) f_0 = _input_f0[_ioff_f0++];
    if (ç\textcolor{blue}{\textit{\_rv2}}ç) s_0 = _input_f0[_ioff_f0++];
    if (ç\textcolor{blue}{\textit{\_rv3}}ç) t_0 = _input_f0[_ioff_f0++];
    if (ç\textcolor{blue}{\textit{\_rv4}}ç ? ((f_0 < s_0) && (f_0 >= t_0)) : ç\textcolor{blue}{\textit{\_ev5}}ç ){
        if (ç\textcolor{blue}{\textit{\_rv6}}ç) _ooff_0 = printInt(_out_0, _ooff_0, f_0);
    }
    if (ç\textcolor{blue}{\textit{\_rv7}}ç ? ((f_0 > s_0) && (s_0 <= t_0)) : ç\textcolor{blue}{\textit{\_ev8}}ç ){
        if (ç\textcolor{blue}{\textit{\_rv9}}ç) _ooff_0 = printInt(_out_0, _ooff_0, s_0);
    }
    if (ç\textcolor{blue}{\textit{\_rv10}}ç? ((f_0 > t_0) && (s_0 > t_0)) : ç\textcolor{blue}{\textit{\_ev11}}ç ){
        if (ç\textcolor{blue}{\textit{\_rv12}}ç) _ooff_0 = printInt(_out_0, _ooff_0, t_0);
    }
    goto scope_1;    
  }
  // scope_1 and scope_2
  final_step:
  assert(strcmp(_out_0, "3") != 0 || ... // other assertions);
}
\end{minted}
\end{code}
\end{figure*}

\subsubsection{\cbmc}
\label{sec:cfaults-cbmc}
After unrolling and instrumentalising the C program, 
\cfaults invokes \cbmc, a bounded model checker for C~\cite{tacas04-cbmc-ClarkeKL}. First, \cfaults invokes \cbmc with memory-safety checks enabled (e.g., \texttt{--bounds-check},\break \texttt{--pointer-check}) to detect inconsistencies in the unrolled, instrumented buggy program that our MBD approach cannot capture, as modelling these violations would cause the SAT solver to consider assignments that do not correspond to valid C states. If \cbmc reports such issues, \cfaults terminates immediately and returns an informative error message to alert the user to invalid memory accesses.
Next, \cfaults uses \cbmc to transform the unrolled and relaxed program into \emph{Static Single Assignment (SSA)} form~\cite{ssa-tpls91}, an intermediate representation ensuring that variables are assigned values only once and are defined before use. SSA achieves this by converting existing variables into multiple versions, each uniquely representing an assignment. Next, \cbmc translates the SSA representation into a CNF formula, which represents the trace formula of the program.
During the CNF formula generation, \cbmc negates the program's assertion ($\neg (\neg A_1 \lor \dots \lor \neg A_m)$) to compute a counterexample. Moreover, the CNF formula, $\phi$, encodes each failing test's input ($I_i$), assertion ($A_i$), and all execution paths of the unrolled and relaxed incorrect program encoded by the trace formula ($P_i$), i.e., $\phi = (I_1\ \land\ \dots\ \land\ I_m)\ \land\ P_i\ \land\ (A_1 \land \dots \land A_m)$. Thus, if $\phi$ is $SAT$, an assignment exists that activates or deactivates each relaxation variable and makes all failing test assertions true. Hence, each satisfiable assignment is a diagnosis of the C program, considering all failing tests. 

\subsubsection{MaxSAT Encoder} 
\label{sec:cfaults-maxsat}
Let $\phi$ denote the CNF formula generated by \cbmc in the previous step. Next, \cfaults generates a weighted partial MaxSAT formula $(\mathcal{H}, \mathcal{S})$ to maximise the satisfaction of relaxation variables in the program, aiming to minimise the necessary code alterations. The set of hard clauses is defined by \cbmc's CNF formula (i.e., $\mathcal{H} = \phi$), while the soft clauses consist of unit clauses representing relaxation variables used to instrument the C program, expressed as $\mathcal{S} = \bigwedge_{c \in \mathcal{C}} { ({rv}_c) }$. 

Additionally, \cfaults assigns hierarchical weights to relaxation variables according to the height of their sub-ASTs (\emph{abstract syntax trees}~\cite{compilers-book-dragon}). For example, for an \texttt{if} statement without an \texttt{else} branch, the relaxation variable associated with the condition receives a weight equal to the sum of the weights of the relaxation variables in the \texttt{then} block. Furthermore, to prioritise identifying faults in the program logic over handling input/output issues, I/O statements (e.g., \texttt{scanf}, \texttt{printf}) are given substantially higher weights than other statements. The hierarchical weighting scheme is optional and can be enabled or configured via \cfaults\ command-line options.

Moreover, 
\cfaults enumerates all \emph{optimal} MaxSAT solutions to obtain every subset-minimal diagnosis of minimum total weight, since multiple optimal solutions (with identical optimum cost) can correspond to different sets of relaxed program statements.

\subsubsection{Oracle} \cfaults invokes a MaxSAT solver to determine the program's diagnoses with minimum weight. Note that all these diagnoses are also subset-minimal, aligning with the principles of Model-Based Diagnosis (MBD) theory. By consolidating all failing tests into a unified, unrolled, and instrumentalised program, the MaxSAT solution identifies the minimum subset of statements requiring removal to fulfil the assertions of all failing tests.

\subsection{\cfaults sees Fault in Circuits}
\label{sec:cfaults-circuits}

As indicated by the orange dashed arrows in Figure~\ref{fig:cfaults-overview}, when \cfaults receives a Boolean circuit $\texttt{B}$ and a test suite as input, it first unrolls the circuit, then instruments the unrolled circuit, and finally encodes it into MaxSAT before invoking a MaxSAT oracle.

\subsubsection{Circuit unrolling}
\label{sec:circuit-unrolling}

\cfaults begins the unrolling process by expanding the faulty Boolean circuit using the set of failing input-output observations from the test suite. 
The unrolled circuit $B_u$ encodes the execution of all failing observations within a single circuit, by encoding their corresponding inputs and outputs. 
During unrolling, \cfaults generates \emph{fresh Boolean variables} for each of the $m$ circuit copies, one for each failing observation, ensuring that every copy has its own unique variables corresponding to its inputs, outputs, and internal gates. 
Thus, an unrolled circuit $B_u$ represents the original circuit \texttt{B} replicated $m$ times, where $m$ is the number of failing test cases, i.e., $B_u \leftrightarrow (i_1  \land B_1 \land o_1) \land (i_2 \land B_2 \land o_2 ) \land  \dots \land  (i_m \land B_m \land o_m)$, where $i_k$ and $o_k$ correspond to the input-output observation of circuit $B_k$.

\subsubsection{Circuit instrumentalisation}

During this step, \cfaults relaxes each inner gate of the unrolled circuit $B_u$, using relaxation variables (i.e., the \emph{healthy variables} described in Section~\ref{sec:mbd-multiple-tests}). 
Each relaxation variable controls the activation of its corresponding circuit component, enabling it when set to \texttt{true} and disabling it when set to \texttt{false}. 
Moreover, identical circuit components (i.e., inner gates) across different circuit copies from different observations share the same relaxation variable. 
As a result, if a relaxation variable is set to \texttt{false}, the corresponding inner gate is deactivated simultaneously across all failing observations. For example, let $g_j$ denote the Boolean variable corresponding to inner gate~$j$. 
Then, \cfaults adds the clause $rv_j \implies g_{jk}$ for each observation $k$, generating the unrolled and relaxed Boolean circuit $B_i$.

\subsubsection{MaxSAT Encoder} 
To encode this MBD problem into MaxSAT, \cfaults generates an unweighted partial MaxSAT formula $(\mathcal{H}, \mathcal{S})$ that maximises the satisfaction of the relaxation variables, thereby minimising the required gate modifications. 
The set of hard clauses is given by the unrolled and instrumented circuit $B_i$, as described in the previous sections (i.e., $\mathcal{H} = B_i$), 
while the soft clauses consist of unit clauses representing the relaxation variables used to instrument the Boolean circuit, expressed as 
$\mathcal{S} = \bigwedge_{c \in \mathcal{C}} ({rv}_c)$, where $\mathcal{C}$ denotes the set of circuit gates.

\subsubsection{Oracle} \cfaults calls a MaxSAT solver to find the circuit's smallest set of faulty components, aligning with the principles of MBD theory with multiple observations presented in Section~\ref{sec:mbd-multiple-tests}. By consolidating all failing input-output examples into a unified and unrolled  circuit, the MaxSAT solution identifies the minimum subset of components requiring removal to fulfil all failing input-output observations.
 
\section{Experimental Results}
\label{sec:results}

All of the experiments were conducted on an Intel(R) Xeon(R) Silver computer with
4110 CPUs @ 2.10GHz running Linux Debian 12, using a memory limit of 32 GB and a timeout of 3600s, for each buggy program, and 1800s for each buggy Boolean circuit. 

Section~\ref{sec:results-c} presents the experiments of running \cfaults and other FBFL methods on benchmarks of C programs. Section~\ref{sec:results-circuits} presents the results of using \cfaults and \hsd~\cite{ijcai19-ignatievMWM} to localise bugs on Boolean circuits.
For the MaxSAT oracle, RC2Stratified~\cite{imms19-RC2} from the \texttt{PySAT} toolkit~\cite{imms18-PySAT}~was~used.

\subsection{C Software}
\label{sec:results-c}

\paragraph{Benchmarks} \cfaults has been evaluated using two distinct benchmarks of C programs: \tcas~\cite{tcas-dataset-ESE05} and \benchmark~\cite{C-Pack-IPAs_apr24}. 

\tcas stands out as a well-known program benchmark extensively studied in the fault localisation literature~\cite{bugAssist-pldi11,jip16-SNIPER}. This benchmark comprises a C program from Siemens and 41 versions with intentionally introduced faults, with known positions and types of these faults. Conversely, \benchmark is a collection of student programs gathered from an introductory programming course. 
For this evaluation, the entire \benchmark benchmark was used, consisting of twenty-five programming assignments and a total of 1431 faulty programs drawn from three laboratory classes. 
These programs cover a range of topics, including integers and input–output operations, loops and character manipulation, and vector processing.
\benchmark has proven successful in evaluating various works across program analysis tasks~\cite{ecai23-GNNs-4-var-mapping,ojm25-aaai,ojm25-jss,OrvalhoJM26}. 

\cfaults uses \texttt{pycparser}~\cite{pycparser}
for unrolling and instrumentalising C programs. Additionally, \cbmc version 5.11 is used to encode C programs into CNF formulae.

\paragraph{Fault Localisation Methods} Since the source code of \bugassist and \sniper is either unavailable or no longer maintained (resulting in compilation and linking issues), prototypes of their algorithms were implemented. It is worth noting that the original version of \sniper could only analyse programs that utilised a subset of ANSI-C, lacked support for loops and recursion, and could only partially handle global variables, arrays, and pointers. Moreover, we have also implemented our prototype for \hsd~\cite{ijcai19-ignatievMWM} to localise faulty statements in ANSI-C software, both with and without core minimisation~(\hsd-CM and \hsd, respectively), since the original algorithm was only implemented for Boolean circuits, and not for C software. Thus, in this work, \bugassist, \sniper and \hsd can handle ANSI-C programs, as their algorithms are built on top of \cfaults's unroller and instrumentaliser~modules. 

All FBFL algorithms evaluated 
consistently generate diagnoses that are consistent with~(\ref{eq:diagnosis-multiple-single-formula}), indicating that all proposed diagnoses undergo validation by \cbmc once the algorithm provides a diagnosis. However, this validation primarily serves to verify diagnoses generated by \bugassist, as it has the capability to produce diagnoses that may not align with all failing test cases. In contrast, both \cfaults' MaxSAT solution, and \hsd hitting set approach, by definition, align with all observations. \sniper's aggregation method (Cartesian product) produces only valid diagnoses, although they may not always be subset-minimal. When considering \bugassist, as presented in Algorithm~\ref{alg:bugassist}, we iterate through all computed diagnoses based on \bugassist's voting score, until we identify one diagnosis that is consistent with all observations, i.e., consistent with (\ref{eq:diagnosis-multiple-single-formula}).

\paragraph{Hierarchical Weights} For both C benchmark, \tcas\ and \benchmark, we evaluate all FBFL approaches \emph{with} and \emph{without} hierarchical weights on the soft clauses (relaxation variables; see Section~\ref{sec:cfaults-maxsat}). 
As the results are similar, we report only the weighted case here; unweighted results appear in~\ref{sec:no-hierarchical-weights}.
As explained in Section~\ref{sec:cfaults-maxsat}, hierarchical weights guide the search toward a minimum set of buggy statements by accounting for the statement’s sub-AST (abstract syntax tree).
Furthermore, since \tcas comprises multiple versions of the same program, the initialisation of program variables can be treated as hard constraints (see Section~\ref{sec:mbd-multiple-tests}), since these initialisations are never faulty in this benchmark. In contrast, \benchmark contains student programs where faults may occur anywhere, including in variable initialisation; accordingly, we relax all executable statements, excluding variable and function declarations, across all FBFL approaches. 

\paragraph{Static Analysers} Before executing any FBFL algorithms, every C program undergoes a series of static checks, as explained in Section~\ref{sec:cfaults}. 
Specifically, prior to unrolling and instrumenting the C programs, the static analysis tools \textsc{cppcheck}~\cite{cppcheck} and \textsc{clang-tidy}~\cite{clang-tidy} are invoked to detect issues such as uninitialised variables, errors like division by zero, and outputs differing only in white space. 
If any issues are detected, the fault localisation procedure is immediately terminated, and the statically identified faults are returned to the user. 
Otherwise, the given buggy program proceeds to unrolling and instrumentation. 
The entire set of programs in the \tcas benchmark was analysed in this way, and no static issues were reported. 
By contrast, the static analysers detected faults such as uninitialised variables in 749 programs from the \benchmark benchmark (approximately 52\%). 
Accordingly, in this section we present FBFL results only for the remaining 48\% of C programs in this~benchmark.

\begin{table}[t!]
\centering
\resizebox{0.8\columnwidth}{!}{%
\begin{tabular}{ccccccc}
\toprule
\multicolumn{7}{c}{Benchmark: \textbf{\tcas}} \\ \hline
{} && \begin{tabular}[c]{@{}c@{}}\textbf{Valid}\\\textbf{Diagnosis}\end{tabular} &&  \textbf{Memouts} && \textbf{Timeouts}\\
\hline
\textbf{\bugassist} & & {\bf 41 (100.0\%)} & & 0 (0.0\%) & & 0 (0.0\%)\\
\textbf{\sniper} & & 11 (26.83\%) & & {\bf 30 (73.17\%)} & & 0 (0.0\%)\\
\textbf{\hsd} & & {\bf 41 (100.0\%)} & & 0 (0.0\%) & & 0 (0.0\%)\\
\textbf{\hsd-CM} & & {\bf 41 (100.0\%)} & & 0 (0.0\%) & & 0 (0.0\%)\\
\textbf{\cfaults} & & {\bf 41 (100.0\%)} & & 0 (0.0\%) & & 0 (0.0\%)\\
\bottomrule
\end{tabular}
}
\caption{Fault localisation results on \tcas~\cite{tcas-dataset-ESE05} benchmark.}
\label{tab:tcas}
\end{table}

\begin{figure*}[t!]
    \begin{subfigure}[t!]{0.49\columnwidth}
    \centering
    \includegraphics[width=1\columnwidth]{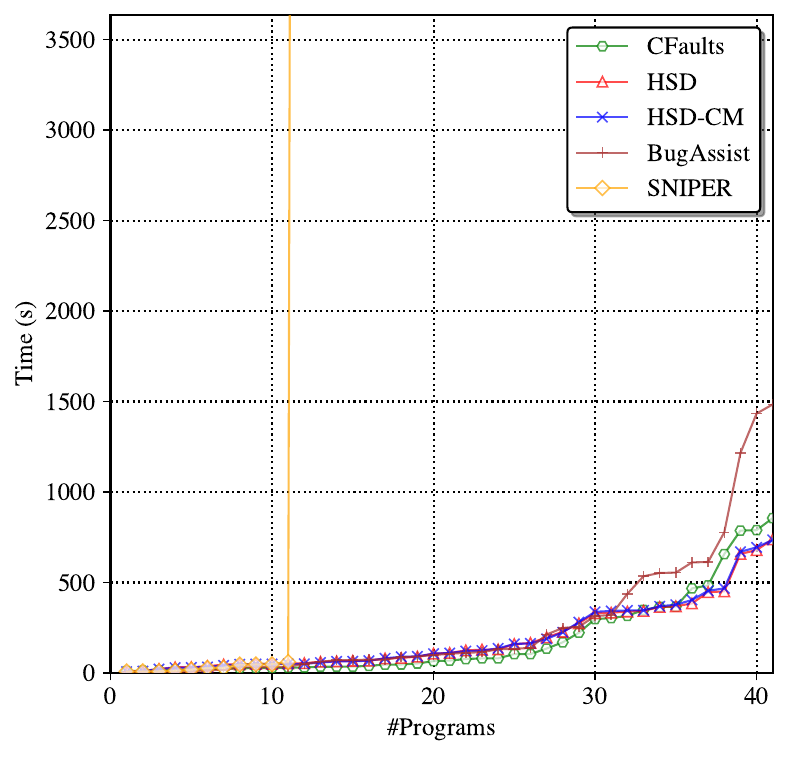} 
     \caption{Time performance of all FBFL approaches.}
     \label{fig:tcas-cpu_time}
     \end{subfigure}
     \hspace{0.02\textwidth}
    \begin{subfigure}[t!]{0.49\columnwidth}
    \centering
    \includegraphics[width=1\columnwidth]{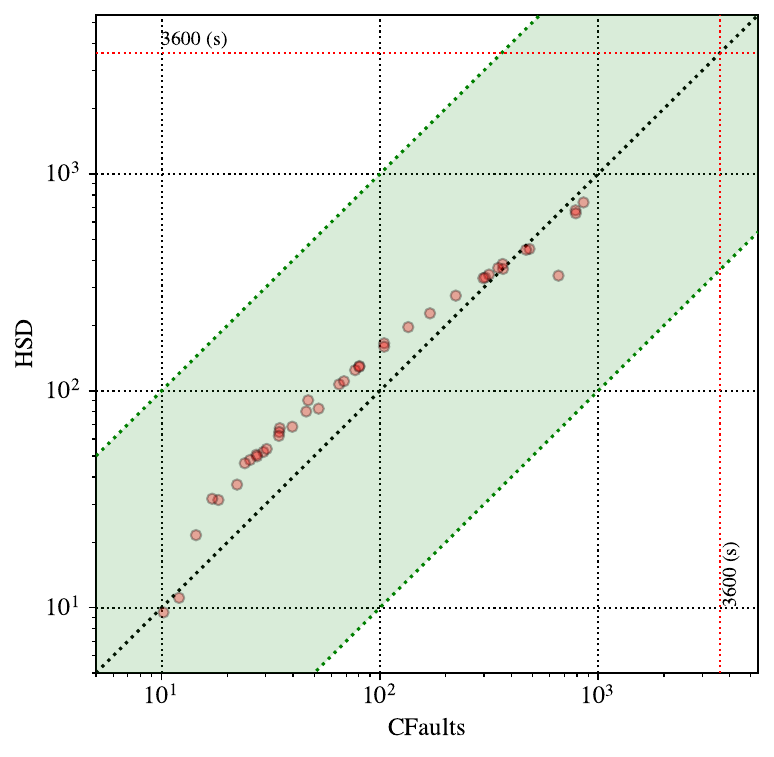}
     \caption{Time comparison between \cfaults and \hsd.}
     \label{fig:tcas-scatter-cpu_time}
     \end{subfigure}
       \caption{Total CPU time comparison between \bugassist, \sniper, \hsd and \cfaults on the \tcas~\cite{tcas-dataset-ESE05} benchmark.}
    \label{fig:tcas-plots} 
\end{figure*}

\paragraph{\tcas} Table~\ref{tab:tcas} provides an overview of the results obtained using \bugassist, \sniper, \hsd, \hsd-CM, and \cfaults on \tcas.  Entries highlighted in bold correspond to the highest value per column. All FBFL methods, except \sniper, find valid diagnoses for every instance in \tcas.
The \tcas program comprises approximately 180 lines of code and has a maximum of 131 failing tests for each program. As shown in Table~\ref{tab:tcas}, this leads \sniper to reach the memory limit of 32GB for 73\% of the programs when aggregating the sets of MCSes computed for each failing test.

Figure~\ref{fig:tcas-cpu_time} presents a \emph{cactus plot} showing the CPU time required for localising faults in each program (y-axis) against the number of programs with faults successfully localised (x-axis), using all FBFL algorithms on \tcas. 
Overall, \cfaults demonstrates faster performance than \bugassist and \sniper, and is generally slightly faster than \hsd. 
The performance of \sniper can be explained by its high memout rate on this benchmark~(see Table~\ref{tab:tcas}). 

Furthermore, Figure~\ref{fig:tcas-scatter-cpu_time} shows a \emph{scatter plot} comparing the CPU times of \cfaults and \hsd on \tcas. 
Each point in the plot corresponds to a faulty program, where the x-value (respectively, y-value) denotes the time spent by \cfaults (respectively, \hsd) in localising faults in that program. 
Points lying above the diagonal indicate programs where \hsd required more CPU time than \cfaults to generate a valid diagnosis. 
As shown, \cfaults is generally faster than \hsd on fault localisation for the \tcas benchmark.

Moreover, although \bugassist localised faults in all programs, it yielded a non-optimal diagnosis in 2.4\% of cases. Additionally, \sniper enumerated redundant diagnoses in 27\% of the programs in which it successfully localised~faults.
 
\begin{table}[t!]
\centering
\resizebox{0.8\columnwidth}{!}{
\begin{tabular}{ccccccc}
\toprule
\multicolumn{7}{c}{Benchmark: \textbf{\benchmark}} \\ \hline
{} && \begin{tabular}[c]{@{}c@{}}\textbf{Valid}\\\textbf{Diagnosis}\end{tabular} &&  \textbf{Memouts} && \textbf{Timeouts}\\
\hline
\textbf{\bugassist} & & 265 (38.86\%) & & 5 (0.73\%) & & 412 (60.41\%)\\ 
\textbf{\sniper} & & 234 (34.31\%) & & {\bf 21 (3.08\%)} & & {\bf 427 (62.61\%)}\\
\textbf{\hsd} & & 375 (54.99\%) & & 12 (1.76\%) & & 295 (43.26\%)\\
\textbf{\hsd-CM} & & 371 (54.40\%) & & 13 (1.91\%) & & 298 (43.70\%)\\
\textbf{\cfaults} & & {\bf 480 (70.38\%)} & & 14 (2.05\%) & & 188 (27.57\%)\\
\bottomrule
\end{tabular}
    }
    \caption{Fault localisation results on \benchmark~\cite{C-Pack-IPAs_apr24} benchmark.}
    \label{tab:c-pack-ipas}
\end{table}

\begin{figure}
    \centering
    \includegraphics[width=0.5\columnwidth]{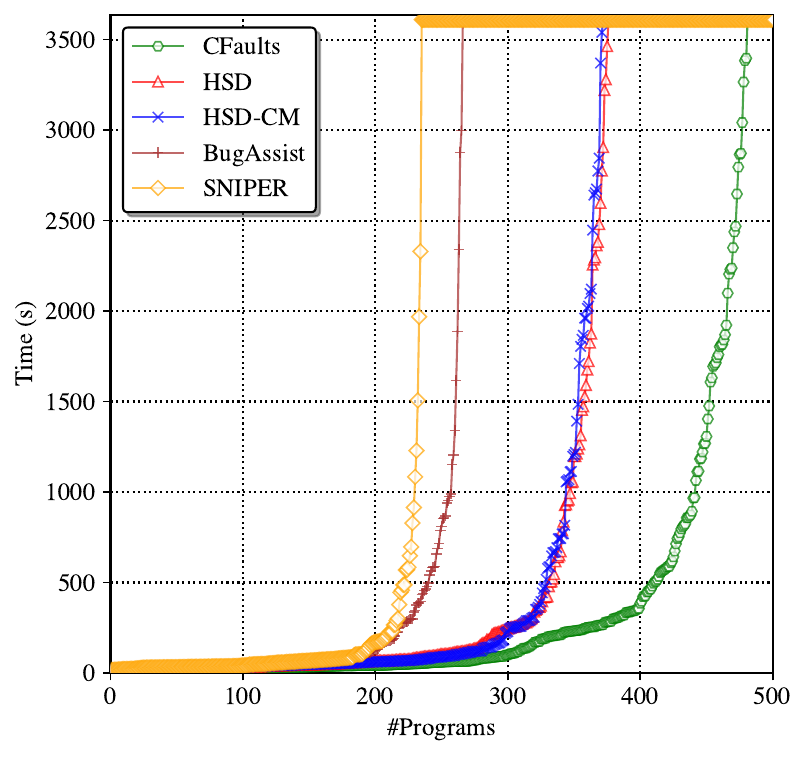}
    \caption{Total CPU time performance of all FBFL approaches on \benchmark.}
    \label{fig:cpackipas-cpu_time}
\end{figure}

\begin{figure*}[t!]
\begin{subfigure}[t!]{0.32\textwidth}    
    \centering
    \includegraphics[width=1\textwidth]{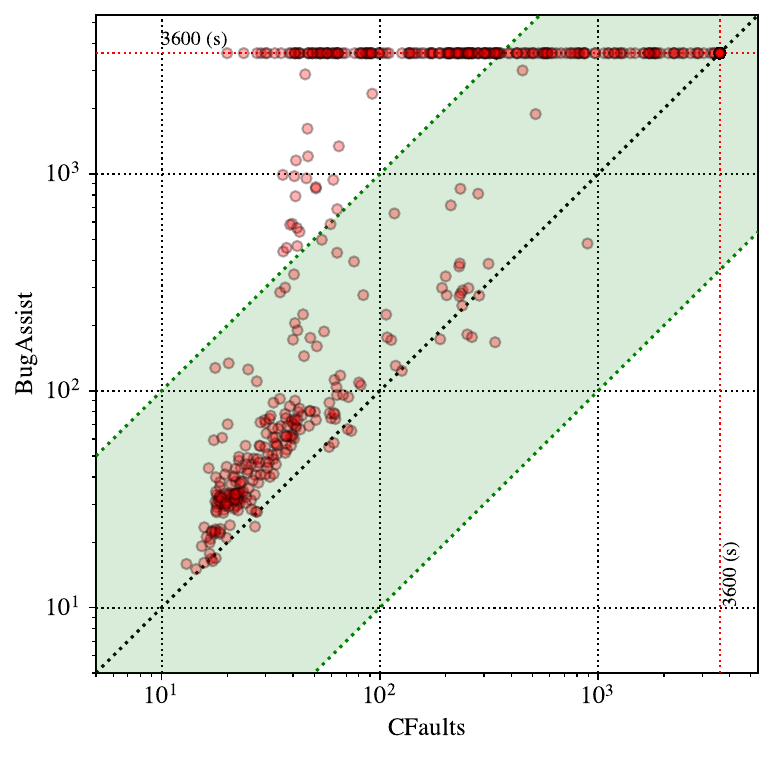}
    \caption{\cfaults vs \bugassist.}
    \label{fig:cpackipas-scatter-cfaults-bugassist}
     \end{subfigure}
     \hspace{0.32\textwidth} 
    \begin{subfigure}[t!]{0.32\textwidth}    
    \centering
    \includegraphics[width=1\textwidth]{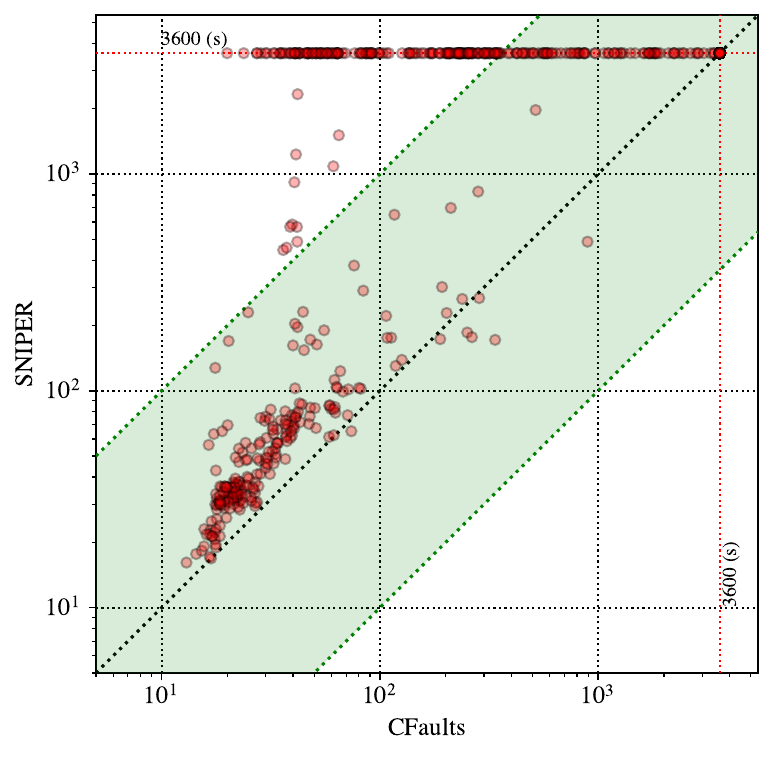}
    \caption{\cfaults vs \sniper.}
    \label{fig:cpackipas-scatter-cfaults-sniper}
     \end{subfigure}\\
    \begin{subfigure}[t!]{0.32\textwidth}    
    \centering
    \includegraphics[width=1\textwidth]{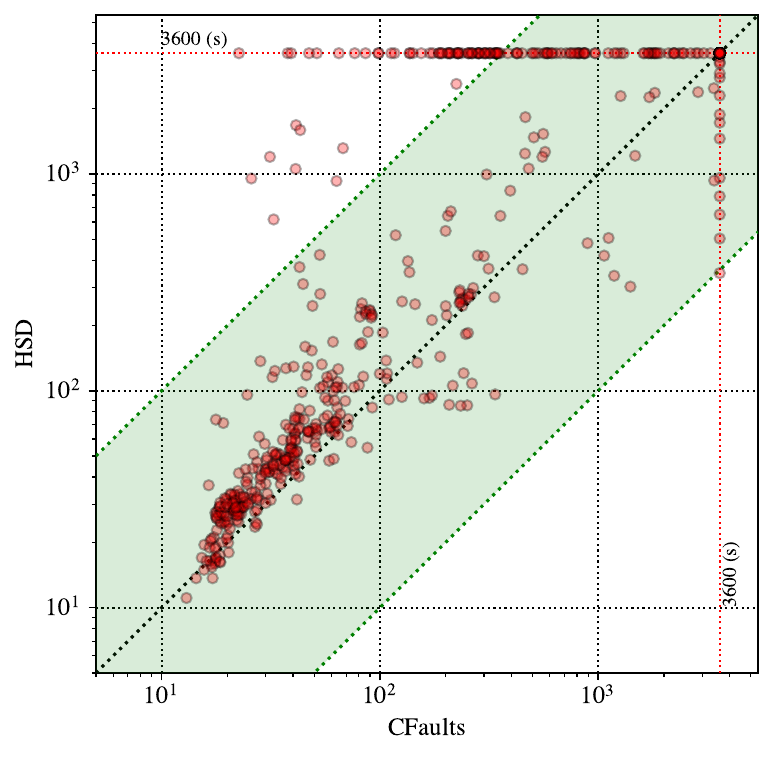}
    \caption{\cfaults vs \hsd.}
    \label{fig:cpackipas-scatter-cfaults-hsd}
     \end{subfigure}
    \begin{subfigure}[t!]{0.32\textwidth}    
    \centering
    \includegraphics[width=1\textwidth]{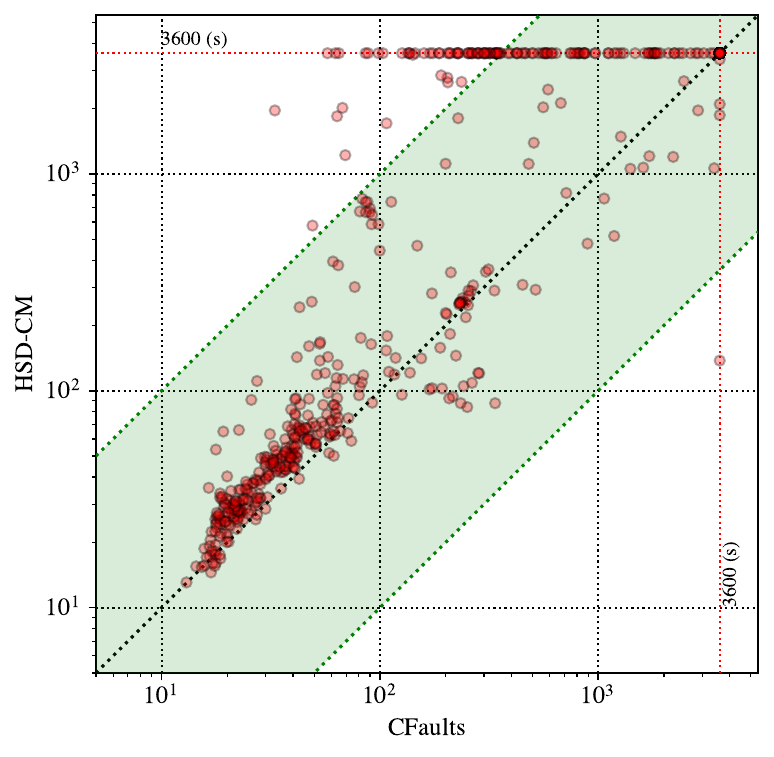}
    \caption{\cfaults vs \hsd-CM.}
    \label{fig:cpackipas-scatter-cfaults-hsd-core-minimisation}
     \end{subfigure}
     \begin{subfigure}[t!]{0.32\textwidth}    
    \centering
    \includegraphics[width=\textwidth]{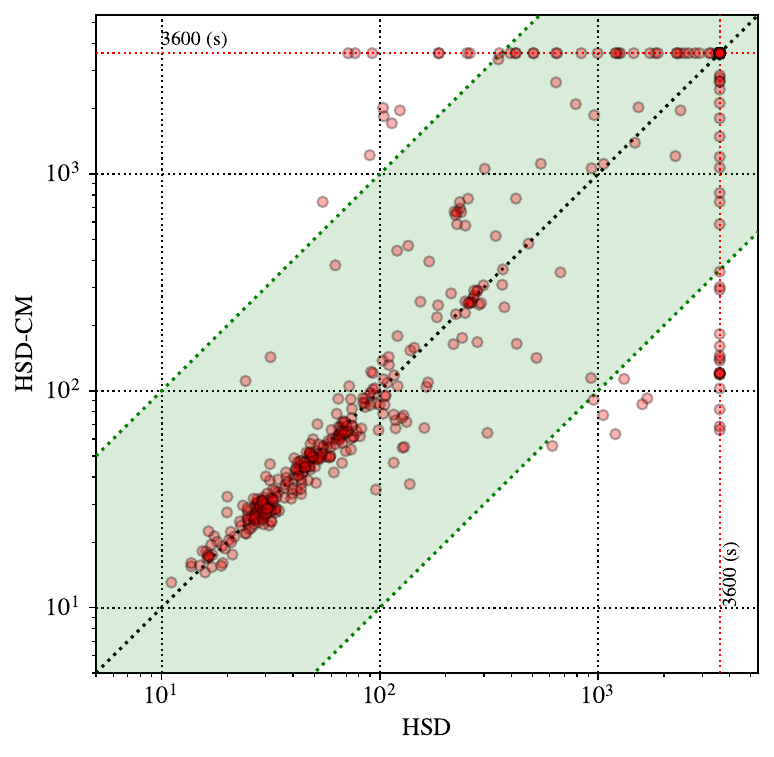}
    \caption{\hsd vs \hsd-CM.}
    \label{fig:cpackipas-scatter-cfaults-hsd-hsd-core-minimisation}
     \end{subfigure}
       \caption{CPU time comparison between \bugassist, \sniper, \hsd, \hsd with core minimisation (\hsd-CM) and \cfaults on \benchmark.}
    \label{fig:cpackipas-scatter-plots-cpu} 
\end{figure*}

\paragraph{\benchmark} Regarding the \benchmark benchmark, Table~\ref{tab:c-pack-ipas} provides an overview of the results obtained using \bugassist, \sniper, \hsd, \hsd-CM, and \cfaults on this benchmark. Entries highlighted in bold correspond to the highest value per column. Figure~\ref{fig:cpackipas-cpu_time} presents a cactus plot of the total CPU time spent by all FBFL algorithms on \benchmark. 
A similar trend can be observed here: \cfaults is faster at localising bugs in C software and is able to localise significantly more bugs than all other FBFL approaches. 
These results are consistent with Table~\ref{tab:c-pack-ipas}, which shows that \cfaults localises faults in 31\% more programs than \bugassist, 36\% more programs than \sniper, and 15\% more programs than \hsd when considering its best-performing configuration on \benchmark (i.e., without core minimisation). 
Moreover, Figure~\ref{fig:cpackipas-scatter-plots-cpu} presents several scatter plots comparing the total CPU time of \cfaults against each of the other localisation methods, further illustrating that, in general, \cfaults is faster than the other FBFL approaches when localising faults in the same~C~programs.

Furthermore, Figure~\ref{fig:cpackipas-scatter-cfaults-hsd-hsd-core-minimisation} shows that the use of unsatisfiable core minimisation in \hsd~(see Section~\ref{sec:hsd}) yields a performance profile similar to that obtained without minimisation. 
For all programs in \benchmark where \hsd successfully localised faults, the minimum, average, and maximum numbers of algorithm iterations (i.e., number of cores enumerated) with core minimisation were 1, 7, and 135, respectively, compared to 1, 105, and 2566 without core minimisation. 
In terms of CPU usage, \hsd-CM devoted on average around 98\% of its total computation time to core minimisation in programs where faults were localised within the time limit. 
For programs where the time limit was exceeded, \hsd-CM spent on average about 93\% of the computation time minimising unsatisfiable cores. 
Thus, while core minimisation enables \hsd to reach minimal diagnoses in fewer iterations, it comes at a cost, as shown in Table~\ref{tab:c-pack-ipas}, the number of faults localised decreases by 0.5\%, memouts increase by 0.15\%, and timeouts increase by around 0.5\%.

\begin{figure*}[t!]
    \begin{subfigure}[t!]{0.48\textwidth}
    \centering
    \centering
    \includegraphics[width=0.8\textwidth]{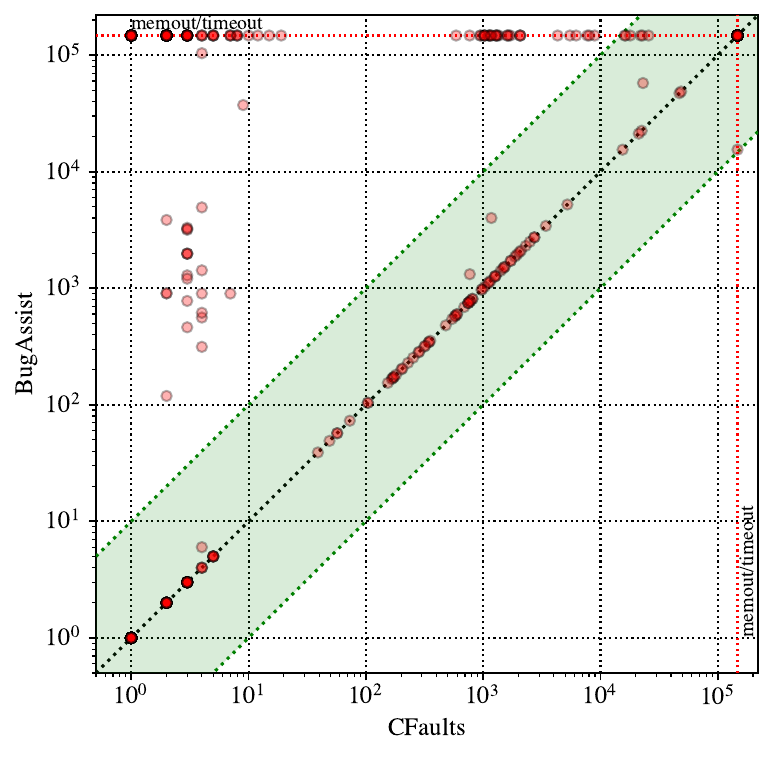}
    \caption{MaxSAT costs of  \bugassist's and \cfaults' diagnoses.}
    \label{fig:scatter-costs}
     \end{subfigure}
     \hspace{0.02\textwidth}
    \begin{subfigure}[t!]{0.48\textwidth}
    \centering
    \includegraphics[width=0.8\textwidth]{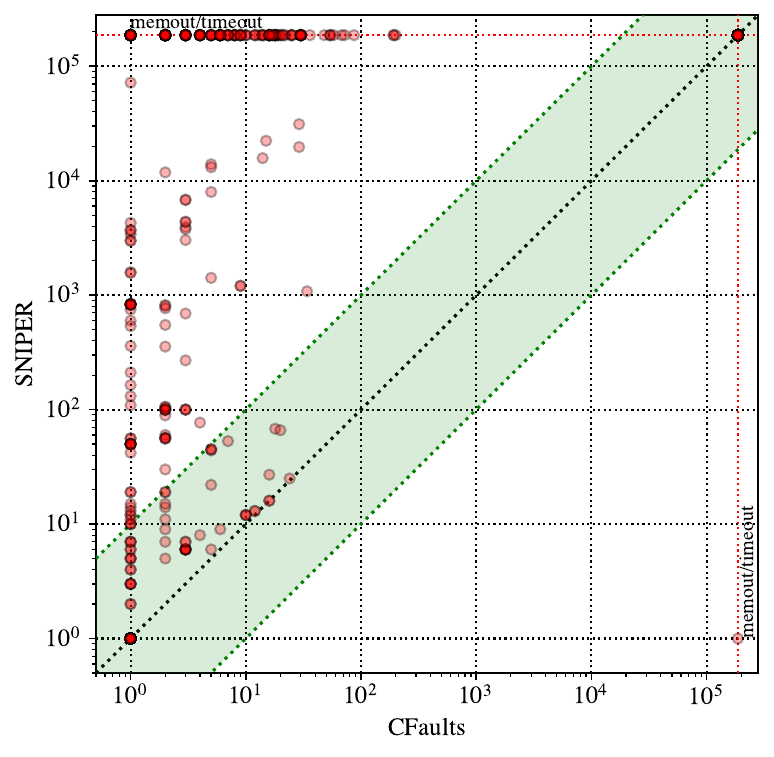}
    \caption{\#Diagnoses generated by \sniper and \cfaults.}
    \label{fig:scatter-num-diagnoses}
    \end{subfigure}
\caption{Comparison between \bugassist's, \sniper's and \cfaults' diagnoses.}
\label{fig:plots}
\end{figure*}

Finally, Figure~\ref{fig:scatter-costs} illustrates a scatter plot comparing the diagnoses' costs of the MaxSAT formulae (i.e., total weight of unsatisfied soft clauses) achieved by \cfaults (x-axis) against \bugassist (y-axis) on \benchmark. 
This plot shows that \bugassist fails to provide an optimal diagnosis in 9\% of cases. 
The other FBFL approaches (i.e., \sniper, \hsd, and \cfaults) always return a minimal diagnosis of optimal cost. However, unlike \cfaults and \hsd, \sniper typically reaches the optimum only after enumerating a large number of redundant (non-optimal) diagnoses.
Figure~\ref{fig:scatter-num-diagnoses} depicts a scatter plot comparing the number of diagnoses enumerated by \cfaults (x-axis) against \sniper (y-axis). This figure shows that while \cfaults and \hsd enumerate all optimal solutions of the weighted MaxSAT formulation, \sniper produces far more diagnoses overall. In particular, \sniper enumerates redundant diagnoses in 55\% of the programs in which it successfully localised faults, suggesting that its enumeration procedure does not prevent redundant diagnoses.
The number of such redundant diagnoses is much larger than the subset-minimal diagnoses generated by \cfaults and \hsd. Figure~\ref{fig:scatter-num-diagnoses} illustrates that in some instances, \sniper may enumerate up to 100K diagnoses, whereas \cfaults generates less than 10 diagnoses.

Thus, although \cfaults incorporates all failing test cases into a single MaxSAT formula and must compute all MaxSAT solutions of that formula,
it remains faster and localises more faults on both C benchmarks, \tcas and \benchmark, than the other FBFL approaches (i.e., \bugassist, \sniper, and \hsd), which divide the localisation process into multiple MaxSAT calls.

\subsection{Boolean Circuits}
\label{sec:results-circuits}

\paragraph{Benchmark} For the Boolean circuit evaluation, we compared \cfaults with the \hsd algorithm~\cite{ijcai19-ignatievMWM}, both with and without core minimisation~(\hsd-CM and \hsd, respectively), on the widely studied \textsc{ISCAS85} benchmark~\cite{iscas85}. 
To extend this benchmark, we injected both single and multiple faults using the publicly available \hsd~\cite{ijcai19-ignatievMWM} code.
In particular, we varied the number of failing observations~(10, 20, 30, 40, and 50) and the number of injected faults~(1, 2, 3, 4, 5, 10, 20, 30,~40, and 50). Thus, the benchmark used in this experiment comprises 12147 buggy Boolean circuits.

For the \iscas benchmark, we evaluate all FBFL approaches \emph{without} hierarchical weights on the soft clauses (i.e., relaxation variables). Such weighting is primarily useful for C programs, where disabling a statement can alter control flow. By contrast, in Boolean circuits, masking a gate does not block the evaluation of downstream gates, so hierarchical weights provide no additional guidance. Hence, for this benchmark we solve an \emph{unweighted} MaxSAT problem, assigning the same cost to all gates.

\begin{table}[t!]
\centering
\resizebox{0.8\columnwidth}{!}{
\begin{tabular}{cccc}
\toprule
\multicolumn{4}{c}{Benchmark: \textbf{\iscas}} \\ \hline
\textbf{\#Observations} & \textbf{CFaults} & \textbf{HSD} & \textbf{HSD-CM}\\\hline
 10 &  2011 (81.62\%) &  \textbf{2135 (86.65\%)} &  2110 (85.63\%) \\
 20 &  1865 (76.12\%) &  \textbf{2004 (81.80\%)} &  1983 (80.94\%) \\
 30 &  1720 (71.70\%) &  \textbf{1878 (78.28\%)} &  1860 (77.53\%) \\
 40 &  1709 (70.82\%) &  \textbf{1886 (78.16\%)} &  1881 (77.95\%) \\
 50 &  1664 (68.73\%) &  1832 (75.67\%) &  \textbf{1844 (76.17\%)} \\ \midrule
 All & 8969 (73.84\%) & \textbf{9735 (80.14\%)} & 9678 (79.67\%) \\
\bottomrule
\end{tabular}
    }
    \caption{Number of successfully localised minimum diagnoses by \hsd (with and without core minimisation) and \cfaults on the augmented \iscas~\cite{iscas85} benchmark. Time limit: 1800s, the memory limit was not reached for any instance.}
    \label{tab:iscas85}
\end{table}

\begin{figure*}[t!]
    \begin{subfigure}[t!]{0.48\textwidth}
    \centering
    \includegraphics[width=0.85\textwidth]{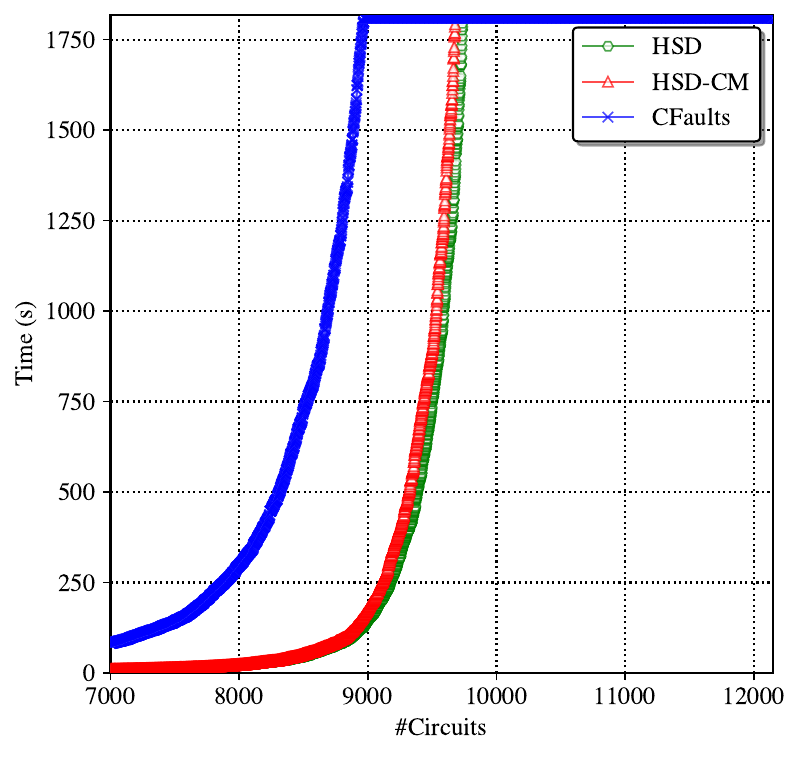}
    \caption{CPU time performance of \hsd and \cfaults.}
    \label{fig:iscas-cpu_time}
     \end{subfigure}
     \hspace{0.02\textwidth}
    \begin{subfigure}[t!]{0.48\textwidth}
    \centering
    \includegraphics[width=0.85\textwidth]{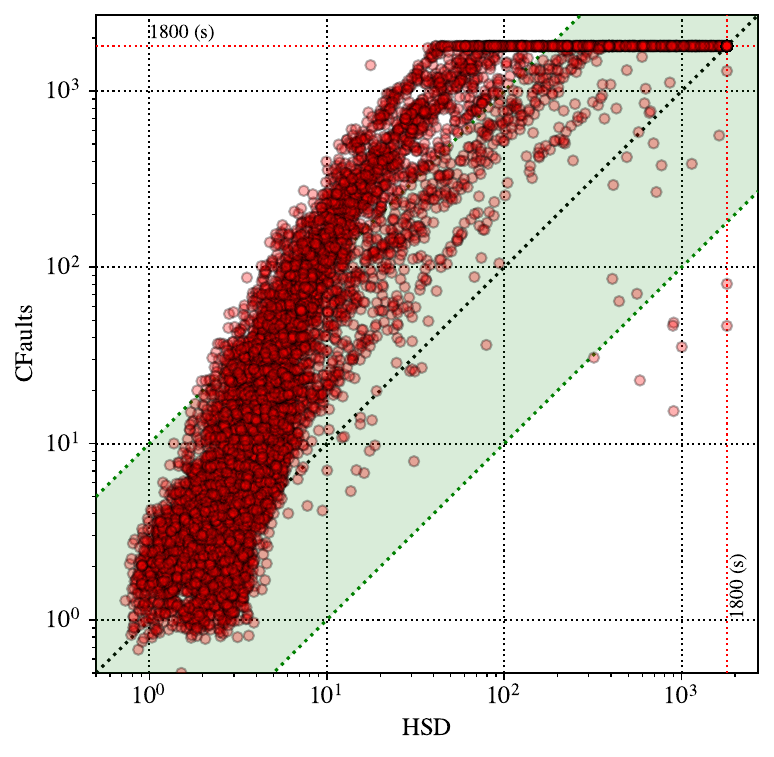}
    \caption{Time comparison between \cfaults and~\hsd.}
    \label{fig:iscas-scatter-time-cfaults-hsd}
    \end{subfigure}
    \begin{subfigure}[t!]{0.48\textwidth}
    \centering
    \includegraphics[width=0.85\textwidth]{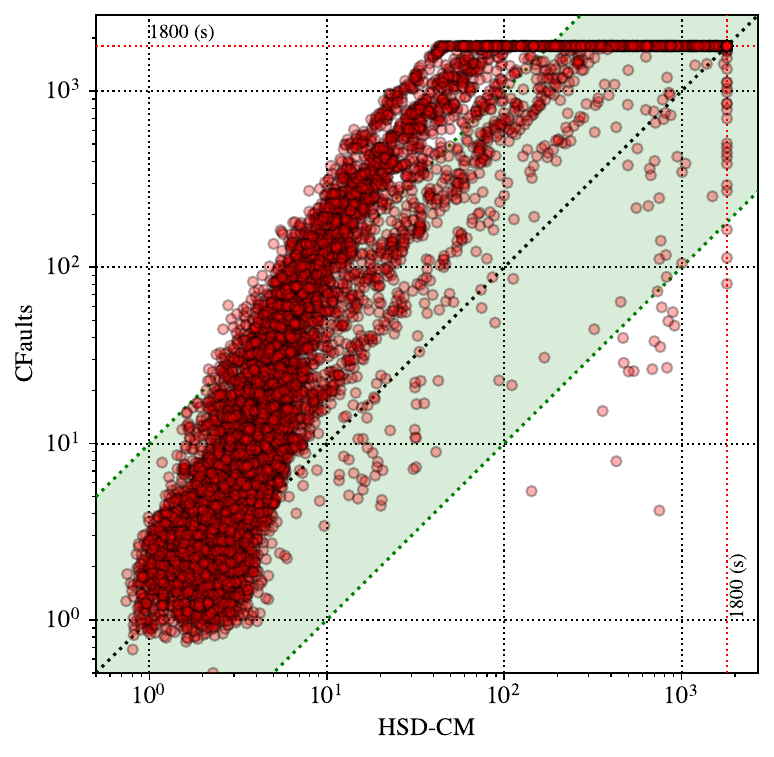}
    \caption{Time comparison between \cfaults and \hsd with core minimisation.}
    \label{fig:iscas-scatter-time-cfaults-hsd-core-minimisation}
    \end{subfigure}
    \begin{subfigure}[t!]{0.48\textwidth}
    \centering
    \includegraphics[width=0.85\textwidth]{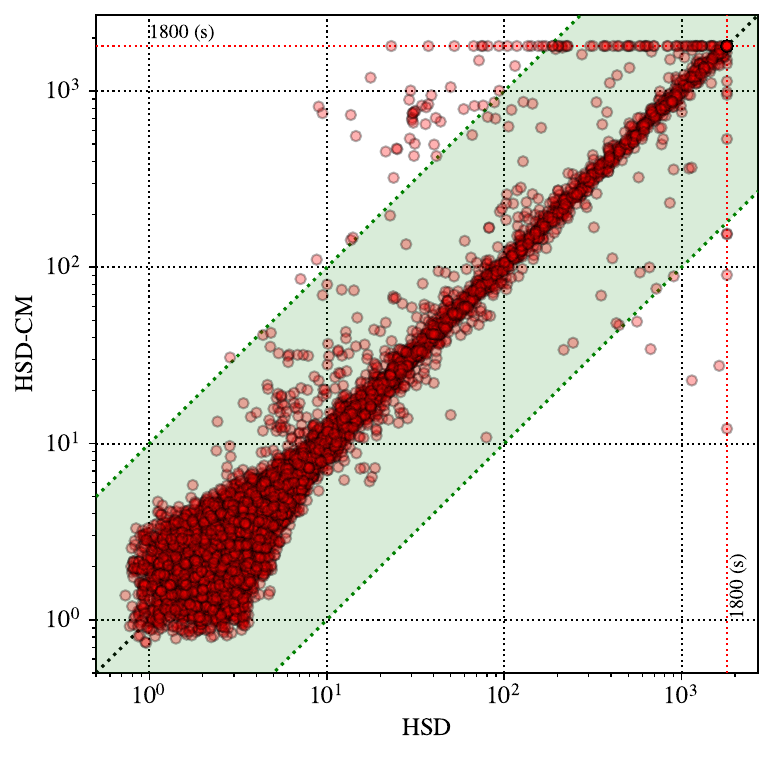}
    \caption{Time comparison between \hsd with and without core minimisation.}
    \label{fig:iscas-scatter-time-hsd-hsd-cm}
    \end{subfigure}
\caption{CPU time comparison between \hsd and \cfaults on the \iscas~\cite{iscas85}.}
\label{fig:iscas-cpu-time-plots}
\end{figure*}

Table~\ref{tab:iscas85} reports the number of minimum diagnoses successfully localised by \hsd, \hsd-CM and \cfaults on the augmented \iscas benchmark, under varying numbers of failing observations. Entries highlighted in bold correspond to the highest value per number of observations.
These results indicate that \cfaults remains highly competitive with \hsd when the number of failing observations is small (10 or 20), localising faults in only 4--5\% fewer circuits than \hsd, depending on whether core minimisation is applied. For larger numbers of failing observations (30, 40, 50), the gap increases modestly to about 6\%. Aggregating over all circuits (i.e., without grouping by observation count), the gap between \cfaults and \hsd is around 6\%. Notably, \hsd performs almost identically with and without core minimisation, differing by~at~most~1\%.

Figure~\ref{fig:iscas-cpu-time-plots} complements these results by comparing the CPU time performance of \cfaults and \hsd. 
In particular, Figure~\ref{fig:iscas-cpu_time} presents a cactus plot depicting CPU time (y-axis) against the number of circuits successfully localised (x-axis), using \cfaults, \hsd, and \hsd-CM on the augmented \iscas benchmark. 
The results confirm that \hsd achieves faster performance overall and localises faults in a larger number of circuits, in line with the findings of Table~\ref{tab:iscas85}. 

Additional insights are provided by the scatter plots in Figures~\ref{fig:iscas-scatter-time-cfaults-hsd},~\ref{fig:iscas-scatter-time-cfaults-hsd-core-minimisation}, and~\ref{fig:iscas-scatter-time-hsd-hsd-cm}, which compare the CPU times of \cfaults and \hsd. 
For example, in Figure~\ref{fig:iscas-scatter-time-cfaults-hsd}, each point corresponds to a faulty circuit, with the x-value~(respectively, y-value) representing the time taken by \hsd~(respectively, \cfaults) to localise faults in that circuit. 
Points lying above the diagonal denote circuits where \cfaults required more time than \hsd to compute a valid diagnosis. 
These plots clearly demonstrate that \hsd, both with and without core minimisation, consistently outperforms \cfaults in terms of CPU~time. 

Moreover, Figure~\ref{fig:iscas-scatter-time-cfaults-hsd-core-minimisation} shows that the use of unsatisfiable core minimisation in \hsd (see Section~\ref{sec:hsd}) results in performance that is almost indistinguishable from that obtained without minimisation. 
For all circuits where \hsd successfully localised faults, the minimum, average, and maximum numbers of cores enumerated without core minimisation were 1, 101, and 33532, respectively, compared to 1, 50, and 6403 with minimisation. 
The average time spent by \hsd on core minimisation in this benchmark was less than 1\%.
For circuits where \cfaults failed to localise faults within the time limit, the corresponding numbers of \hsd iterations were 4, 543, and 33532 without core minimisation, and 4, 155, and 6012 with minimisation.

In summary, on the augmented \iscas benchmark with multiple injected faults and varying numbers of failing observations, \cfaults is generally slower and succeeds in localising faults in fewer Boolean circuits overall than \hsd. Nevertheless, \cfaults remains competitive, localising faults in only 6\% fewer circuits than \hsd.

\subsection{Discussion}

In the evaluation on buggy C software, using the \tcas and \benchmark benchmarks, \cfaults demonstrated clear advantages over the competing FBFL methods. 
Even though \cfaults encodes all failing test cases into a single MaxSAT formula,
it consistently outperformed the other approaches. 
In particular, \cfaults was both faster and able to localise more faults on \benchmark, identifying faults in 31\% more programs than \bugassist, 36\% more programs than \sniper, and 15\% more programs than \hsd, all of which decompose the fault localisation task into multiple independent MaxSAT~problems.

By contrast, in the evaluation on Boolean circuits with multiple injected faults and varying numbers of failing observations, the performance trends differ. 
In this setting, \cfaults is generally slower than \hsd and localises faults in fewer circuits overall. 
However, the difference is modest since \cfaults localises faults in only 6\% fewer circuits than \hsd, indicating that it remains a competitive approach.

Taken together, these results suggest that among the FBFL approaches considered, \cfaults is the fastest and most reliable method for C software, whereas for Boolean circuits it remains competitive, but \hsd continues to represent the state of the art. 
One factor helps to explain this difference in performance. 
Fault localisation in C programs and Boolean circuits constitutes two structurally distinct problems, leading to MaxSAT encodings with significantly different formulae structures. 
This distinction has a substantial influence on solver efficiency, contributing to the observed disparity between the two domains.

\section{Threats to Validity}
\label{sec:threats}

\paragraph{Faulty code not being executed}
Our approach requires that faulty statements are \emph{executed} by at least one failing test case; statements that are never executed cannot be diagnosed. Consequently, insufficient input coverage can yield false negatives (i.e., missed diagnoses). We partially mitigate this with multiple failing observations and lightweight static analysis.

\paragraph{Bounded model checking (BMC)}
We rely on bounded encodings of programs. Loop unrolling and depth bounds can exclude feasible behaviours, which may render otherwise valid diagnoses unreachable under a given bound. Moreover, when bounds truncate iterations, diagnoses that hold for the bounded model may not generalise to the original program. We choose bounds conservatively and report them alongside results, but incompleteness due to bounding is an inherent limitation.

\paragraph{Pointers, arrays, and undefined behaviour}
As explained in Section~\ref{sec:cfaults-progs},\break \cfaults relies on static analysers such as, \textsc{cppcheck}~\cite{cppcheck} and \textsc{clang-tidy}~\cite{clang-tidy}, to detect issues such as uninitialised variables and potential run-time errors (e.g., division by zero). If any such issues are detected, \cfaults terminates immediately and reports them to the user. Furthermore, \cfaults enables \cbmc's memory-safety checks (e.g., \texttt{--bounds-check},\break \texttt{--pointer-check}) to detect invalid memory operations, including out-of-bounds accesses and unsafe pointer arithmetic. However, \cfaults inherits \cbmc's specification and encoding of ANSI~C: behaviours deemed well defined by \cbmc's memory model are treated as valid by our encoding, whereas violations that \cbmc does not flag may escape detection. Clarifying, and, where possible, tightening, the alignment between \cbmc's configuration and the intended C semantics for pointers and arrays is left for future work.

\paragraph{External validity}
Our evaluation uses benchmark and student programs that compile successfully and include explicit specifications (assertions or expected outputs). Generalising to larger, system-scale codebases; programs with extensive I/O, concurrency, or libraries with complex contracts; or to specifications expressed in richer formalisms may require additional engineering and could alter the results observed in this study.

\section{\cfaults: Use Cases}
\label{sec:use-cases}

This section presents several use cases of fault localisation in C software where \cfaults has been successfully applied.

\emph{Providing feedback in programming tasks.}  
In Spring 2024, we evaluated the use of \cfaults to provide automated error feedback to students in a first-year undergraduate programming course in C~\cite{ojm24-sigcse}. 
\cfaults was integrated with \textsc{GitSEED}~\cite{ojm24-sigcse}, a \textsc{GitLab}-backed automated assessment tool for software engineering and programming education, which served as the platform for assessing laboratory work and projects in the course. 
Among the findings of this study, 69\% of the surveyed students reported that the ``hints'' (i.e., localised faults) generated by \cfaults were helpful. 
These hints provided valuable guidance by pointing students towards potential errors in their code and promoting a deeper understanding of programming concepts through self-directed correction. 
Overall, this study~\cite{ojm24-sigcse} highlighted the positive impact of \cfaults as a feedback mechanism, demonstrating its value as a comprehensive educational tool in programming courses.

\emph{Guiding LLMs on automated program repair.} \cfaults has also been integrated with Large Language Models~(LLMs), via zero-shot learning, to enhance automated program repair for C programming assignments~\cite{ojm25-aaai,OrvalhoJM26}. 
In this setting, \cfaults identifies the buggy parts of a C program and presents the LLM with a program sketch in which these faulty statements are removed. 
This hybrid approach follows a Counterexample Guided Inductive Synthesis~(CEGIS) loop~\cite{cegis} to iteratively refine the program. 
The LLM is then asked to synthesise the missing parts, which are subsequently checked against a test suite. 
If the synthesised program is incorrect, a counterexample from the test suite is provided to the LLM to guide a revised synthesis. 
This method enabled LLMs to repair more programs and produce smaller fixes, outperforming both alternative configurations that do not employ FBFL methods and state-of-the-art symbolic program repair tools.

\emph{Formula-based fault localisation for Python.}  
Some bugs in Python can be detected by transpiling code to C with LLMs and applying \cfaults to the generated C code~\cite{ok25-arXiv-PyVeritas}. 
\textsc{PyVeritas} is a framework that performs high-level transpilation from Python to C, followed by bounded model checking and MaxSAT-based fault localisation through \cfaults. 
This approach enables verification and fault localisation for Python programs using existing model-checking tools for C. 
Empirical evaluation on two Python benchmarks shows this method can reach accuracies of up to 80--90\% for some~LLMs.

\section{Related Work}
\label{sec:related}

Fault localisation (FL) techniques typically fall into two main families:
\emph{spectrum-based (SBFL)} and \emph{formula-based (FBFL)}. 
SBFL methods~\cite{DBLP:conf/kbse/AbreuZG09,DBLP:journals/jss/WongDC10,DBLP:journals/tosem/NaishLR11,DBLP:journals/tr/WongDGL14,DBLP:journals/tse/WongGLAW16,DBLP:journals/jss/AbreuZGG09} 
estimate the likelihood of a component being faulty based on test coverage information collected from both passing and failing test executions. 
While SBFL techniques are generally fast, they may lack precision, as not all components identified in this way are necessarily responsible for the observed failures~\cite{liu2019you,DBLP:conf/cav/RothenbergG20}.

In contrast, FBFL approaches~\cite{DBLP:journals/ai/KleerW87,DBLP:conf/aaai/MetodiSKC12,bugAssist-cav11,CFaults-FM24,DBLP:conf/aaai/SternKFP12,DBLP:journals/jair/MetodiSKC14,ijcai15-Marques-SilvaJI15,ijcai19-ignatievMWM,bugAssist-pldi11,jip16-SNIPER,DBLP:journals/ws/ShchekotykhinFFR12,DBLP:conf/dx/RodlerS17,DBLP:journals/kbs/Rodler22,jannach2010toward,DBLP:journals/ai/RodlerHJNW25,lambda2,DBLP:conf/icfem/LamraouiN14,DBLP:journals/entcs/GriesmayerSB07,DBLP:journals/jlp/WotawaNM12,DBLP:conf/popl/XieA05,DBLP:conf/fmcad/KonighoferB11,DBLP:conf/aaai/ZhouOZ022} 
are considered exact. 
As explained in Section~\ref{sec:mbd}, these methods (e.g., \bugassist~\cite{bugAssist-pldi11}, \sniper~\cite{jip16-SNIPER}) deal with the problem of fault localisation as a \emph{Model-Based Diagnosis (MBD)}~\cite{reiter87} problem, seeking minimal sets of faulty components. Typically, they make a MaxSAT call for each failing test case, yielding a minimal diagnosis for each failing observation in isolation rather than reasoning over all failures jointly. They then apply an aggregation procedure to aggregate the per-test diagnoses into a single explanation of the system’s inconsistency. As we show in this paper, for C software such formula-based fault localisation (FBFL) methods either produce a large number of redundant diagnoses (e.g., \sniper) or fail to return a minimal set sufficient to repair the program (e.g., \bugassist).
Recently, \citet{graussam2024consistency} proposed a model-based software fault localisation approach for ANSI~C, introducing a well-defined fault model that covers complex language features (including pointers and arrays) and building on \cbmc encodings; consequently, the relaxation step is performed at the \cbmc level rather than at the code level, as in our work. The approach also handles multiple failing observations by formulating standard MBD instances and solving them with \hsd~\cite{ijcai19-ignatievMWM} to guarantee minimal diagnoses per failing test case. However, \citet{graussam2024consistency} applies an aggregation procedure after diagnosis enumeration that may compromise the minimality of the final diagnoses. We did not include this tool in our evaluation for two main reasons. First, in its current form, functions to be analysed must be manually annotated, an impractical requirement at our scale (we analyse over 1500 programs). Second, the tool currently supports only integers and floats in C, which makes it unsuitable for analysing the \benchmark benchmark.

Finally, Model-Based Diagnosis has been successfully applied to restore consistency in several domains, including Boolean circuits~\cite{DBLP:journals/ai/KleerW87,DBLP:conf/aaai/MetodiSKC12,DBLP:conf/aaai/SternKFP12,DBLP:journals/jair/MetodiSKC14,ijcai15-Marques-SilvaJI15,ijcai19-ignatievMWM,DBLP:conf/aaai/ZhouOZ022}, 
C software~\cite{CFaults-FM24,bugAssist-pldi11,jip16-SNIPER,phd2025mentor}, 
knowledge bases~\cite{DBLP:journals/ws/ShchekotykhinFFR12,DBLP:conf/dx/RodlerS17,DBLP:journals/kbs/Rodler22}, 
and spreadsheets~\cite{jannach2010toward,DBLP:journals/ai/RodlerHJNW25}. 
\emph{Program slicing}~\cite{DBLP:journals/ese/SoremekunKBZ21,DBLP:conf/cav/RothenbergG20,zeller1999yesterday} 
has also emerged as a complementary technique for fault localisation in programs. 
A more syntactic fault localisation approach~\cite{DBLP:conf/cav/RothenbergG20} employs program slicing to enumerate all minimal sets of repairs for a given faulty program. 
Another method for identifying the causes of faulty behaviour involves analysing the differences between successive software~versions~\cite{zeller1999yesterday}. 

\section{Conclusion}
\label{sec:conclusion}

This paper introduces a novel formula-based fault localisation technique for C programs and Boolean circuits, capable of handling any number of faults. 
By leveraging Model-Based Diagnosis (MBD) with multiple observations, \cfaults consolidates all failing test cases into a single MaxSAT formula, thereby ensuring consistency throughout the localisation process and avoiding redundant diagnoses. 

Experimental evaluations highlight the effectiveness of this unified approach. 
On C software benchmarks, \tcas and \benchmark, \cfaults consistently outperforms existing FBFL methods such as \bugassist,\ \sniper, and \hsd. 
In particular, \cfaults was both faster and able to localise more faults on \benchmark, identifying faults in 31\% more programs than \bugassist, 36\% more programs than \sniper, and 15\% more programs~than~\hsd. 

In contrast, on the \iscas benchmark of Boolean circuits with multiple injected faults, \cfaults is generally slower than \hsd and localises faults in slightly fewer circuits overall. 
However, the gap is modest since \cfaults localises faults in only 6\% fewer circuits than \hsd, demonstrating that it remains competitive in this domain.

Overall, these results establish \cfaults as the fastest and most reliable FBFL approach for C software, while for Boolean circuits it remains competitive with the state-of-the-art, represented by \hsd. 
This work thus provides the first unified framework for formula-based fault localisation across both C software and Boolean circuits domains.


\section*{Acknowledgements}
PO and MK acknowledge support from the ERC under the European Union’s Horizon 2020 research and innovation programme (FUN2MODEL, grant agreement No.~834115) and ELSA: European Lighthouse on Secure and Safe AI project (grant agreement
No. 101070617 under UK guarantee).
This work was partially supported by Portuguese national funds through FCT, under projects UID/50021/2025, UID/PRR/50021/2025, PTDC/\-CCI-COM/\-2156/2021 (DOI: 10.\-54499/\-PTDC/\-CCI-COM/\-2156/\-2021), 2023.14280.\-PEX (DOI: 10.54499/\-2023.14280.PEX) and 2024.07127.CBM, and grant SFRH/\-BD/\-07724/\-2020 (DOI: 10.54499/\-2020.\-07724.BD).
This work was also supported by the MEYS within the program ERC CZ under the project POSTMAN no.~LL1902 and co-funded by the EU under the project \emph{ROBOPROX} (reg.~no.~CZ\-.02.01.01/00/\-22\_008/0004590).

\bibliographystyle{elsarticle-num-names} 
\bibliography{mybibliography}







\newpage

\appendix

\section{Experiments on C Software Without Hierarchical Weights}
\label{sec:no-hierarchical-weights}

In this section, we evaluate all FBFL approaches on both C benchmarks, \tcas~\cite{tcas-dataset-ESE05} and \benchmark~\cite{C-Pack-IPAs_apr24}, \emph{without} hierarchical weights on the soft clauses (relaxation variables; see Section~\ref{sec:cfaults-maxsat}). Results \emph{with} hierarchical weights are reported in Section~\ref{sec:results-c}. 

\subsection{\tcas}
\label{sec:tcas-no-hierarchical-weights}

\begin{table}[b!]
\centering
\resizebox{0.8\columnwidth}{!}{%
\begin{tabular}{ccccccc}
\toprule
\multicolumn{7}{c}{Benchmark: \textbf{\tcas}} \\ \hline
{} && \begin{tabular}[c]{@{}c@{}}\textbf{Valid}\\\textbf{Diagnosis}\end{tabular} &&  \textbf{Memouts} && \textbf{Timeouts}\\
\hline
\textbf{\bugassist} & & {\bf 41 (100.0\%)} & & 0 (0.0\%) & & 0 (0.0\%)\\
\textbf{\sniper} & & 11 (26.83\%) & & {\bf 30 (73.17\%)} & & 0 (0.0\%)\\
\textbf{\hsd} & & {\bf 41 (100.0\%)} & & 0 (0.0\%) & & 0 (0.0\%)\\
\textbf{\hsd-CM} & & {\bf 41 (100.0\%)} & & 0 (0.0\%) & & 0 (0.0\%)\\
\textbf{\cfaults} & & {\bf 41 (100.0\%)} & & 0 (0.0\%) & & 0 (0.0\%)\\
\bottomrule
\end{tabular}
}
\caption{\bugassist, \sniper, \hsd and \cfaults fault localisation results on \tcas~\cite{tcas-dataset-ESE05} benchmark, not using hierarchical weights.}
\label{tab:tcas-unweighted}
\end{table}

\begin{figure*}[b!]
    \begin{subfigure}[t!]{0.49\columnwidth}
    \centering
    \includegraphics[width=0.8\columnwidth]{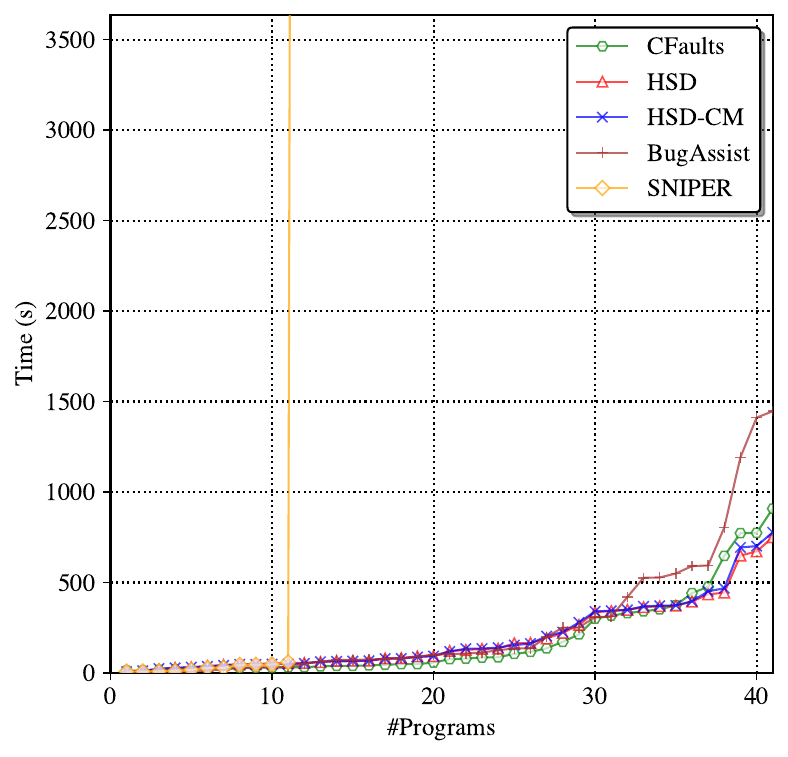} 
     \caption{Time performance of all FBFL approaches.}
     \label{fig:tcas-cpu_time-unweighted}
     \end{subfigure}
     \hspace{0.02\textwidth}
    \begin{subfigure}[t!]{0.49\columnwidth}
    \centering
    \includegraphics[width=0.8\columnwidth]{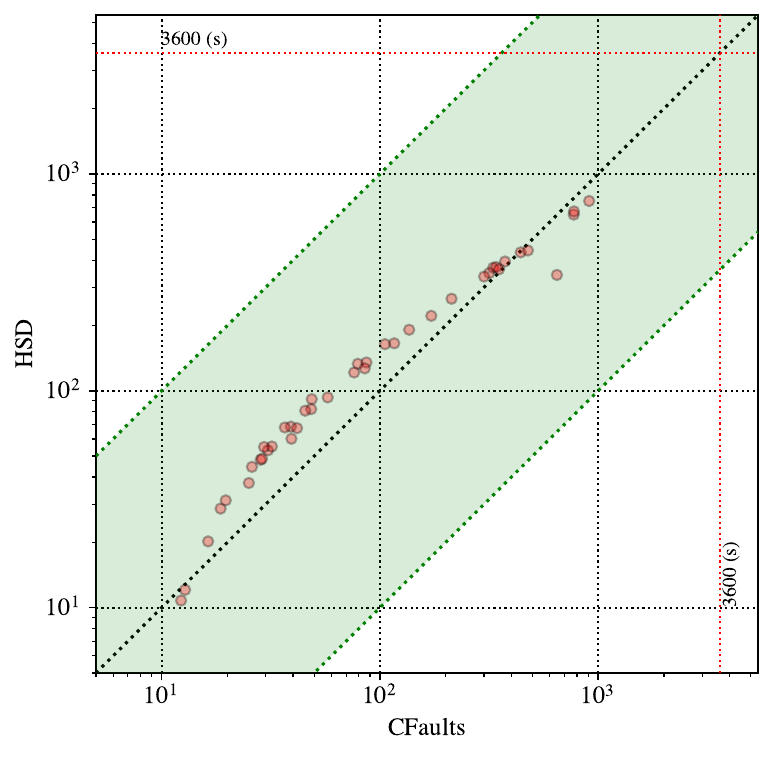}
     \caption{Time comparison between \cfaults and \hsd.}
     \label{fig:tcas-scatter-cpu_time-unweighted}
     \end{subfigure}
       \caption{Total CPU time comparison between \bugassist, \sniper, \hsd and \cfaults on the \tcas~\cite{tcas-dataset-ESE05} benchmark, not using hierarchical weights.}
    \label{fig:tcas-plots-unweighted} 
\end{figure*}

Table~\ref{tab:tcas-unweighted} summarises the results for \bugassist, \sniper, \hsd, \hsd-CM, and \cfaults on \tcas \emph{without} hierarchical weights.  Entries highlighted in bold correspond to the highest value per column. The outcomes are similar to those obtained \emph{with} hierarchical weights (see Table~\ref{tab:tcas}), all FBFL methods, except \sniper, find valid diagnoses for every instance in \tcas.

Figure~\ref{fig:tcas-cpu_time-unweighted} shows a \emph{cactus plot} of CPU time (y-axis) versus the number of successfully localised faulty programs (x-axis) for all FBFL algorithms on \tcas without hierarchical weights. \cfaults follows the same trend as in the hierarchical weighted setting: it is faster than \bugassist and \sniper, and typically slightly faster than \hsd.
Figure~\ref{fig:tcas-scatter-cpu_time-unweighted} presents a \emph{scatter plot} comparing \cfaults and \hsd CPU times on \tcas\ without hierarchical weights. Each point corresponds to a faulty program; the x\mbox{-}coordinate (resp.\ y\mbox{-}coordinate) is the time taken by \cfaults\ (resp.\ \hsd) to localise the fault. Points above the diagonal indicate cases where \hsd\ required more time than \cfaults. As shown, \cfaults\ is generally faster than \hsd\ on \tcas\ even in the unweighted~setting.

\subsection{\benchmark}
\label{sec:cpackipas-no-hierarchical-weights}

\begin{table}[t!]
\centering
\resizebox{0.8\columnwidth}{!}{
\begin{tabular}{ccccccc}
\toprule
\multicolumn{7}{c}{Benchmark: \textbf{\benchmark}} \\ \hline
{} && \begin{tabular}[c]{@{}c@{}}\textbf{Valid}\\\textbf{Diagnosis}\end{tabular} &&  \textbf{Memouts} && \textbf{Timeouts}\\
\hline
\textbf{\bugassist} & & 284 (41.64\%) & & 4 (0.59\%) & & 394 (57.77\%)\\ 
\textbf{\sniper} & & 248 (36.36\%) & & {\bf 21 (3.08\%)} & & {\bf 413 (60.56\%)}\\
\textbf{\hsd} & & 382 (56.01\%) & & 11 (1.61\%) & & 289 (42.38\%)\\
\textbf{\hsd-CM} & & 369 (54.11\%) & & 12 (1.76\%) & & 301 (44.13\%)\\
\textbf{\cfaults} & & {\bf 469 (68.77\%)} & & 15 (2.20\%) & & 198 (29.03\%)\\
\bottomrule
\end{tabular}
    }
    \caption{\bugassist, \sniper, \hsd and \cfaults fault localisation results on \benchmark~\cite{C-Pack-IPAs_apr24} benchmark, without hierarchical weights.}
    \label{tab:c-pack-ipas-unweighted}
\end{table}

\begin{figure}[t!]
    \centering
    \includegraphics[width=0.4\columnwidth]{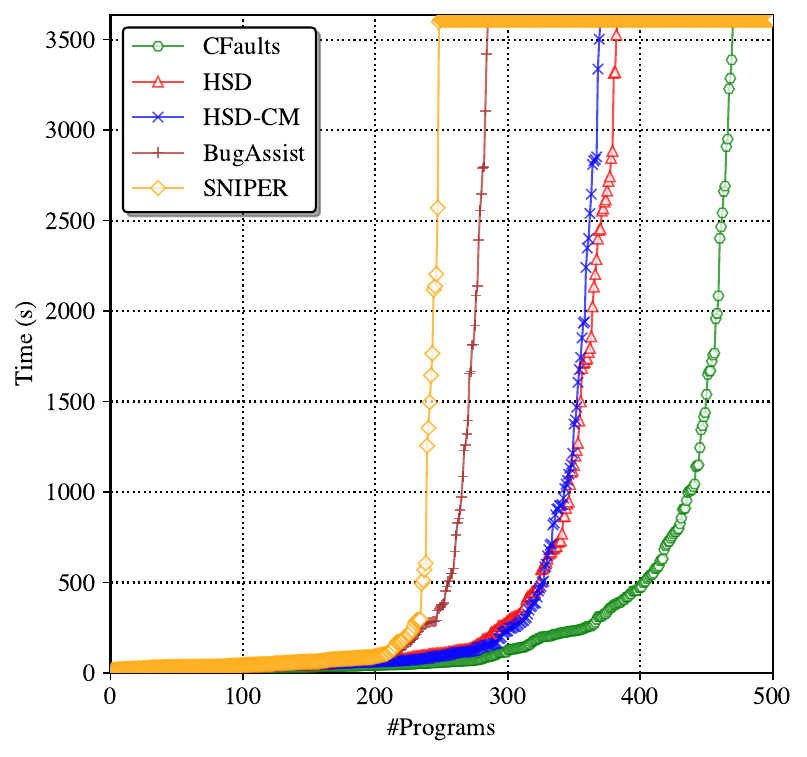}
    \caption{Total CPU time performance of all FBFL approaches on \benchmark, not using hierarchical weights.}
    \label{fig:cpackipas-cpu_time-unweighted}
\end{figure}

\begin{figure*}[b!]
\begin{subfigure}[t!]{0.32\textwidth}    
    \centering
    \includegraphics[width=1\textwidth]{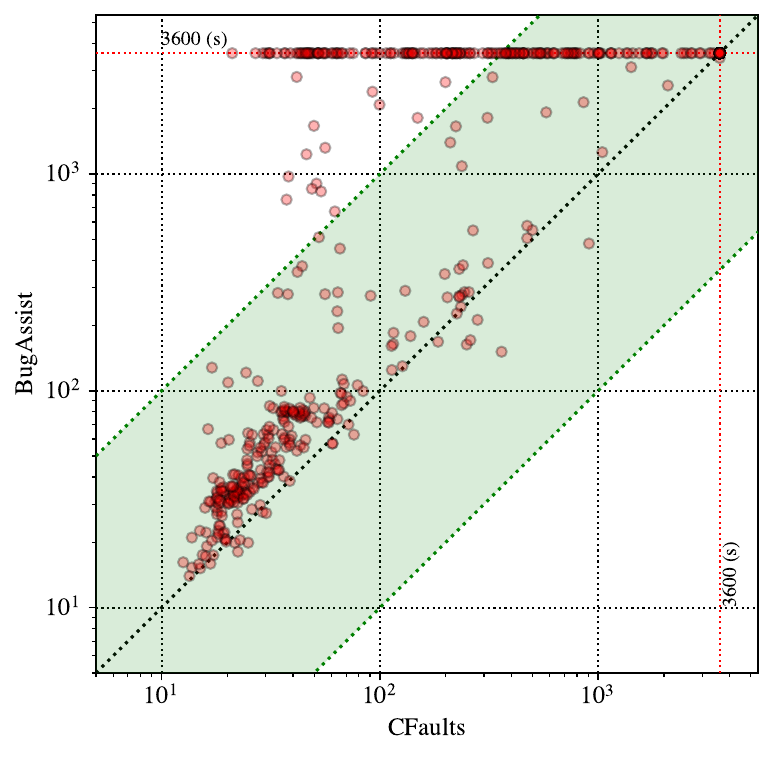}
    \caption{\cfaults vs \bugassist.}
    \label{fig:cpackipas-scatter-cfaults-bugassist-unweighted}
     \end{subfigure}
     \hspace{0.32\textwidth} 
    \begin{subfigure}[t!]{0.32\textwidth}    
    \centering
    \includegraphics[width=1\textwidth]{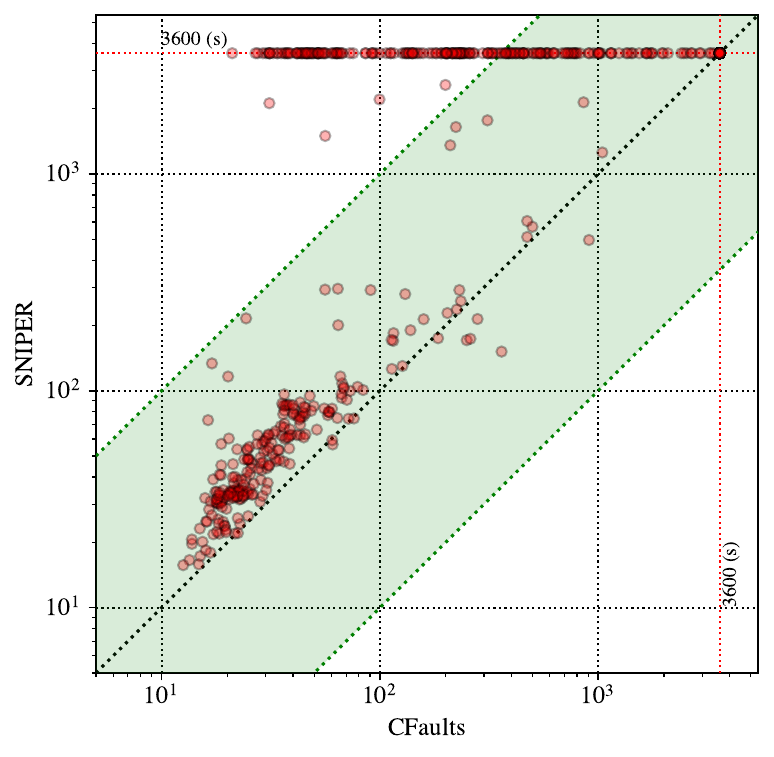}
    \caption{\cfaults vs \sniper.}
    \label{fig:cpackipas-scatter-cfaults-sniper-unweighted}
     \end{subfigure}\\
    \begin{subfigure}[t!]{0.32\textwidth}    
    \centering
    \includegraphics[width=1\textwidth]{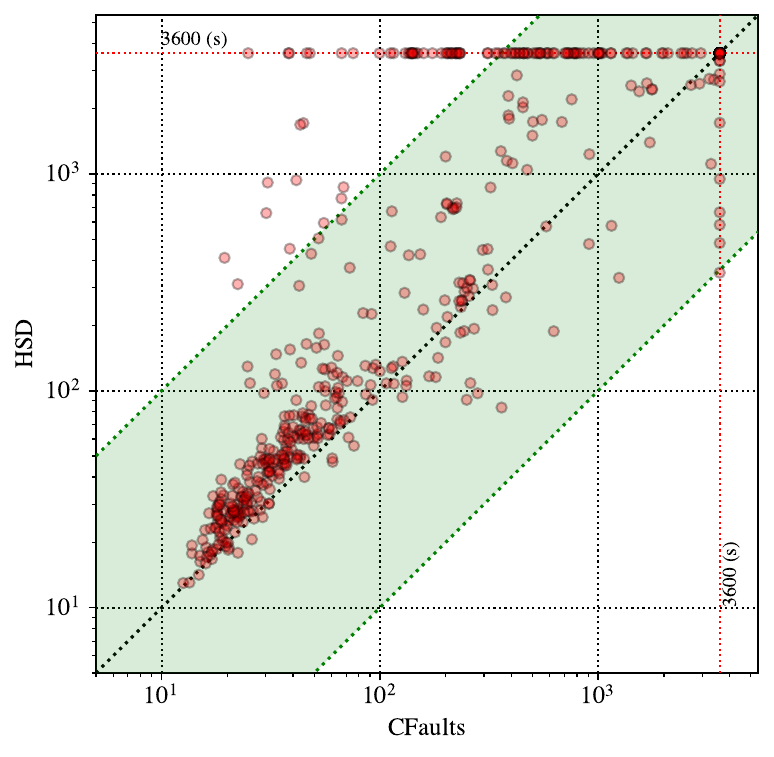}
    \caption{\cfaults vs \hsd.}
    \label{fig:cpackipas-scatter-cfaults-hsd-unweighted}
     \end{subfigure}
    \begin{subfigure}[t!]{0.32\textwidth}    
    \centering
    \includegraphics[width=1\textwidth]{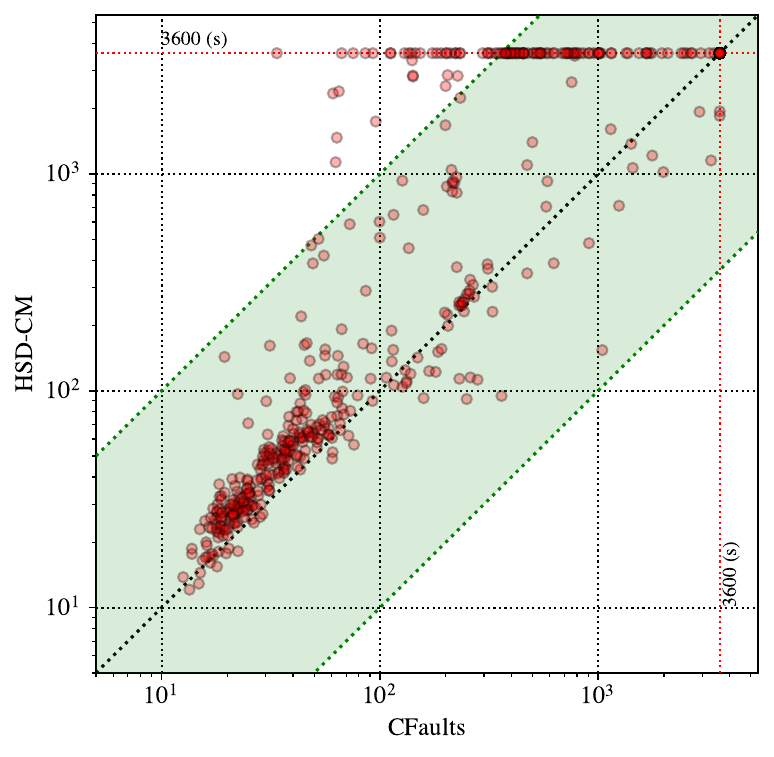}
    \caption{\cfaults vs \hsd-CM.}
    \label{fig:cpackipas-scatter-cfaults-hsd-core-minimisation-unweighted}
     \end{subfigure}
     \begin{subfigure}[t!]{0.32\textwidth}    
    \centering
    \includegraphics[width=\textwidth]{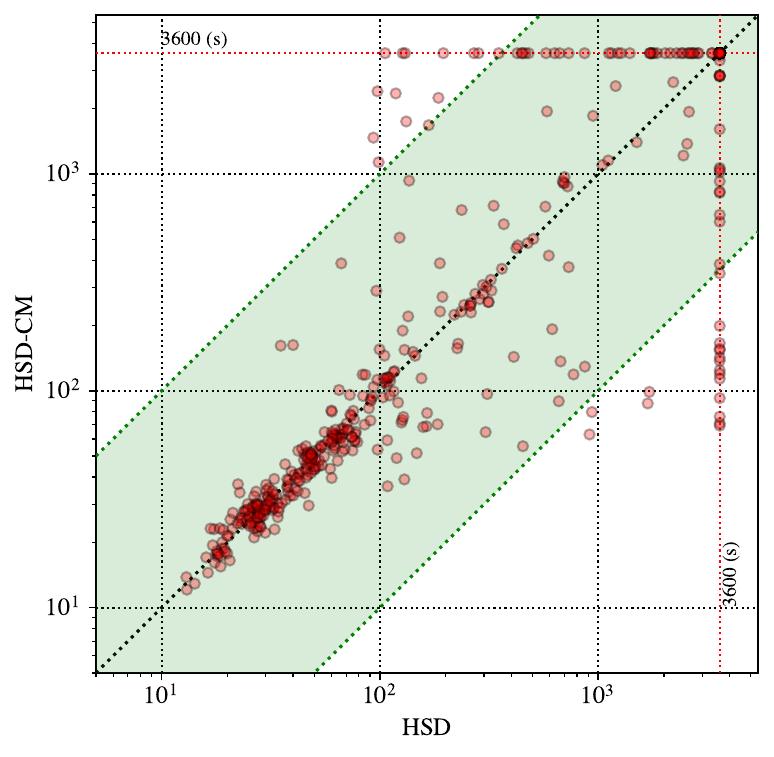}
    \caption{\hsd vs \hsd-CM.}
    \label{fig:cpackipas-scatter-cfaults-hsd-hsd-core-minimisation-unweighted}
     \end{subfigure}
       \caption{CPU time comparison between \bugassist, \sniper, \hsd, \hsd with core minimisation (\hsd-CM) and \cfaults on \benchmark, not using hierarchical weights.}
    \label{fig:cpackipas-scatter-plots-cpu-unweighted} 
\end{figure*}

Table~\ref{tab:c-pack-ipas-unweighted} summarises results for \bugassist, \sniper, \hsd, \hsd\text{-}CM, and \cfaults\ on \benchmark\ \emph{without} hierarchical weights. Entries highlighted in bold correspond to the highest value per column. Figure~\ref{fig:cpackipas-cpu_time-unweighted} provides a cactus plot of total CPU time across all FBFL algorithms.  Without hierarchical weights, \cfaults\ localises faults in 28\% more programs than \bugassist, 33\% more than \sniper, and 13\% more than \hsd\ (using \hsd's best configuration on \benchmark, i.e., without core minimisation). These outcomes are similar to those obtained \emph{with} hierarchical weights (see Table~\ref{tab:tcas}).
Figure~\ref{fig:cpackipas-scatter-plots-cpu-unweighted} presents scatter plots comparing \cfaults\ with each competing method in terms of total CPU time, further showing that, in general, \cfaults\ is faster than the other FBFL approaches when localising faults in the same C programs without hierarchical weights.

\end{document}